\documentclass[11pt]{article}
\newfont{\bbbold}{msbm10 scaled \magstep1}

\def\bbR{\mbox{\bbbold R}}

\def\cE{{\cal E}}
\def\cF{{\cal F}}
\def\cG{{\cal G}}

\def\cL{{\cal L}}

\newfont{\goth}{eufm10 scaled \magstep1}

\def\gi{\mbox{\goth i}}

\def\gn{\mbox{\goth n}}
\def\go{\mbox{\goth o}}
\def\gp{\mbox{\goth p}}

\def\gs{\mbox{\goth s}}

\def\a{\alpha}
\def\b{\beta}
\def\c{\gamma}\def\C{\Gamma}
\def\d{\delta}
\def\e{\epsilon}
\def\f{\phi}\def\F{\Phi}
\def\h{\eta}

\def\k{\kappa}
\def\l{\lambda}\def\L{\Lambda}
\def\m{\mu}
\def\n{\nu}

\def\r{\rho}
\def\s{\sigma}\def\S{\Sigma}
\def\t{\tau}
\def\th{\theta}

\def\beq{\begin{equation}}\def\eeq{\end{equation}}
\def\beqa{\begin{eqnarray}}\def\eeqa{\end{eqnarray}}
\def\barr{\begin{array}}\def\earr{\end{array}}

\def\o{\omega}\def\O{\Omega}
\def\del{\partial}
\def\ua{\underline{\alpha}}
\def\ub{\underline{\phantom{\alpha}}\!\!\!\beta}
\def\uc{\underline{\phantom{\alpha}}\!\!\!\gamma}

\def\um{\underline{\mu}}
\def\ud{\underline\delta}

\def\uo{\underline\omega}\def\uO{\underline{\Omega}}
\def\una{\underline a}\def\unA{\underline A}
\def\unb{\underline b}\def\unB{\underline B}
\def\unc{\underline c}\def\unC{\underline C}
\def\und{\underline d}
\def\une{\underline e}\def\unE{\underline E}
\def\unf{\underline{\phantom{e}}\!\!\!\! f}\def\unF{\underline F}
\def\ung{\underline g}\def\unG{\underline G}
\def\unm{\underline m}\def\unM{\underline M}\def\unK{\underline K}
\def\unn{\underline n}\def\unN{\underline N}
\def\unp{\underline{\phantom{a}}\!\!\! p}\def\unP{\underline P}

\def\uny{\underline{y}}
\def\unH{\underline{H}}
\def\unF{\underline{F}}\def\unT{\underline{T}}\def\unR{\underline{R}}
\def\unK{\underline{K}}

\def\xz{\times}

\def\nab{\nabla}
\def\tT{\tilde{T}}\def\tR{\tilde{R}}\def\tO{\tilde{\Omega}}\def\tE{\tilde{E}}
\def\hE{\hat{E}}\def\hT{\hat{T}}\def\hR{\hat{R}}\def\hO{\hat{\Omega}}
\def\iE{(E^{-1})}\def\nno{\nonumber}
\def\tnab{\tilde{\nabla}}
\def\hM{\hat M}\def\hG{\hat G}\def\hW{\hat W}
\def\hu{\hat u}
\def\huM{\hat{\unM}}\def\huG{\hat{\unG}}
\def\huA{\hat{\unA}}\def\huB{\hat{\unB}}\def\huC{\hat{\unC}}

\def\huo{\hat{\uo}}
\def\huK{\hat{\unK}}\def\ho{\hat{\omega}}
\def\tE{\tilde{E}}

\def\tf{\tilde{f}}
\let\la=\label

\def\um{{\underline m}}

\def\nn{\nonumber}
\def\bd{\begin{document}}
\def\ed{\end{document}}
\def\ba{\begin{array}}
\def\ea{\end{array}}
\def\bea{\begin{eqnarray}}
\def\eea{\end{eqnarray}}
\def\ft#1#2{{\textstyle{{\scriptstyle #1}\over {\scriptstyle #2}}}}
\def\fft#1#2{{#1 \over #2}}
\newcommand{\be}{\begin{equation}}
\newcommand{\ee}{\end{equation}}
\newcommand{\eq}[1]{(\ref{#1})}
\newcommand{\w}[1]{\\[0.#1cm]}
\def\eqs#1#2{(\ref{#1}-\ref{#2})}
\def\det{{\rm det\,}}
\def\tr{{\rm tr}}
\newcommand{\hoch}[1]{$\, ^{#1}$}
\newcommand{\tamphys}{\it\small Center for Theoretical Physics,
Texas A\&M University, College Station, TX 77843, USA}
\newcommand{\kings}
{\it\small Department of Mathematics, King's College, London, UK}
\makeatletter
\renewcommand\theequation{\thesection.\arabic{equation}}
\@addtoreset{equation}{section} \makeatother

\newcommand{\auth}{\large P.S. Howe\hoch{1} and E. Sezgin \hoch{2}}

\begin{document}

\hfill{KCL-TH-04-16}

\hfill{MIFP-04-28/04}

\hfill{hep-th/0412245}

\hfill{\today}

\vspace{20pt}

\begin{center}
{\Large{\bf The supermembrane revisited}} \vspace{30pt}

\auth

\vspace{15pt}

\begin{itemize}
\item [$^1$] \kings \item [$^2$] \tamphys
\end{itemize}

\vspace{60pt}

{\bf Abstract}

\end{center}

The M2-brane is studied from the perspective of superembeddings. We
review the derivation of the M2-brane dynamics and the supergravity
constraints from the standard superembedding constraint and we discuss
explicitly the induced $d=3, N=8$ superconformal geometry on the
worldvolume. We show that the gauged supermembrane, for a target space
with a  $U(1)$ isometry, is the standard D2-brane in a type IIA
supergravity background. In particular, the D2-brane action, complete
with the Dirac-Born-Infeld term, arises from the gauged Wess-Zumino
worldvolume 4-form via the brane action principle. The discussion is
extended to the massive D2-brane considered as a gauged supermembrane
in a massive $D=11$ superspace background. Type IIA supergeometry is
derived using Kaluza-Klein techniques in superspace.

\pagebreak \tableofcontents \setcounter{page}{1}

\section{Introduction}


The supermembrane \cite{Bergshoeff:1987cm} is an important ingredient of
M-theory. It arises as a solution of eleven-dimensional supergravity
\cite{Duff:1990xz}, the low-energy limit of M-theory, and is one of the
candidates for an underlying microscopic definition of the theory, at
least in a particular background via the matrix model
\cite{deWit:1988ig,Banks:1996vh}. The type IIA string arises from the
supermembrane via double dimensional reduction \cite{Duff:1987bx}, while
vertical reduction leads to the effective action for the D2-brane
\cite{Duff:1992hu,Townsend:1995af}.

A membrane traces out a 3-dimensional worldvolume as it moves
through spacetime, but the supermembrane also has fermionic degrees
of freedom and can be viewed as a bosonic extended object moving
through superspace. This gives rise to a Green-Schwarz type action
which is a spacetime supersymmetric version of the Nambu-Goto action
together with a Wess-Zumino term. The ratio between the two terms in
the action is fixed, as in the Green-Schwarz superstring
\cite{Green:1983wt}, by the requirement that the action be invariant
under a local fermionic symmetry \cite{Bergshoeff:1987cm}  called
kappa-symmetry \cite{Siegel:1983hh}. This symmetry implies that the
fully gauge-fixed worldvolume theory is supersymmetric. It was
realised some time ago, first in the case of superparticles
\cite{Sorokin:1989zi,Sorokin:1988nj} moving in low spacetime
dimensions and subsequently for other extended objects, reviewed in
\cite{Sorokin:1999jx}, that kappa-symmetry can be re-interpreted as
local worldvolume supersymmetry. This is particularly clear from the
point of view of superembeddings where the local supersymmetry on
the bosonic worldvolume can be identified with the leading component
( in a $\th$ expansion) of an odd super-worldvolume diffeomorphism.
The supermembrane itself was discussed from this point of view in
\cite{Bandos:1995zw} (see \cite{Sorokin:1999jx} for a review). The
superembedding formalism can be applied to all BPS branes, including
D-branes \cite{Howe:1996mx}; in particular, it was used to obtain
the dynamics of the M5-brane in a covariant formalism
\cite{Howe:1996yn,Howe:1997fb}. The dynamics of the M5-brane were
subsequently derived in a suitably modified Green-Schwarz formalism
\cite{Pasti:1997gx,Aganagic:1997zq}.

There is a universal constraint which is imposed in the superembedding
formalism which, in many cases, determines the worldvolume dynamics,
the constraints on the target space supergeometry and the induced
geometry on the worldvolume \cite{Howe:1996mx,Howe:1997wf}. This is
true for M-branes and most D-branes; all D-branes can be accommodated
in the formalism if open strings ending on them are taken into account
\cite{Chu:1998jv}. This approach leads naturally to the equations of
motion, but an action principle has also been developed in the
superembedding formalism \cite{Bandos:1995dw,Howe:1998ts}. Moreover,
the basic embedding constraint is closely related to the characteristic
form of kappa-symmetry transformations. The superembedding formalism
also provides a convenient framework to study higher-derivative
corrections to the target space supergeometry or to the worldvolume
dynamics. This requires the embedding constraint to be modified as was
shown in the case of the M2-brane in \cite{Howe:2003sa}, and more
recently for the M5-brane in \cite{Drummond:2004cw}. Higher-derivative
corrections for the D9-brane, where the basic embedding condition does
not suffice to give the lowest-order dynamics, have also been
constructed in the superembedding formalism \cite{Drummond:2004vf}.

In this paper we focus on the lowest order dynamics of the M2-brane in
the superembedding framework. We begin by reviewing the superembedding
formalism from two points of view, with and without introducing a
connection on the target superspace. We show how the embedding constraint
requires the constraints on the target space supergeometry to be those
that describe on-shell $D=11$ supergravity. We then provide the full
details of the derivation of the M2-brane dynamics, including the
equations of motion in a supergravity background in component form. We
review the brane action principle and use it to derive the Green-Schwarz
action for the supermembrane starting from a closed 4-form on the
worldvolume. We show how one can determine the worldvolume supervielbein
and give expressions for the bosonic worldvolume dreibein and the
worldvolume gravitino as composite fields. We discuss the structure of
the induced $d=3, N=8$ supergeometry and show that it coincides with that
of off-shell $d=3, N=8$ conformal supergravity.

In section 6 we consider the gauged supermembrane. This has been
discussed in components in \cite{Ortin:1997jh,Bergshoeff:1997ak}, and in
superspace in \cite{Nishino:2003yn}, but has not been discussed previously in
the superembedding formalism. It is shown that the gauged superembedding
formalism leads directly to the D2-brane in IIA superspace without the need to
dualise the additional scalar coming from the eleventh direction. This is done
at the level of the equations of motion via the natural gauged version of the
embedding constraint; it is also shown how  the D2-brane action can be derived
using an appropriate modification of the construction of the Wess-Zumino
term in gauged sigma models. In section 7 we study the massive D2-brane
\cite{Bergshoeff:1997cf} starting from $D=11$
\cite{Ortin:1997jh,Bergshoeff:1997ak}. The $D=11$ supergeometry
corresponding to the lifted massive type IIA supergravity is described.
This is not truly eleven-dimensional since it relies on the presence of a
Killing vector, but it facilitates the construction of the brane theory
via gauged superembedding. As in the massless case one finds that the
gauged superembedding formalism leads directly to the equations of motion
of the D2-brane. The action can also be constructed. As expected, it
includes a Chern-Simons term for the Born-Infeld gauge field proportional
to the mass.

Our conventions are summarised in Appendix A. In Appendix B we give an
of the basic features of embeddings in Riemannian geometry, and in
Appendix C we obtain the superspace constraints of type IIA
supergravity by Kaluza-Klein reduction from eleven-dimensional
superspace.


\section{Superembeddings: connection-free formalism}


\subsection{The embedding constraint}


In ordinary Lorentzian geometry the basic object is the metric tensor
$g$ which can be brought to the standard form $\h_{ab}$ in an
orthonormal basis. If $M$ is an embedded submanifold of $\unM$, the
metric $\underline{g}$ on $\unM$ induces a metric $g$ on $M$ in the
standard way. However, in supergeometry, the basic object is not
$\h_{ab}$ but rather the Dirac matrices $(\C^c)_{\a\b}$. This gives
rise to a three-index object but it is not  a tensor on the whole
space. In order to facilitate the discussion of superembeddings it is
therefore useful to formalise this structure. We shall do it for the
case of $D=11$ but the discussion can be generalised to arbitrary
superspaces. However, not only is $D=11$ the case of interest for us it
is also particularly simple as we shall see.

Let $\unM$ be an $(11|32)$-dimensional (real) supermanifold. We suppose
that we are given a choice of odd tangent bundle $\unF\subset \unT$,
where $\unT$ is the tangent bundle. We can define a three-index tensor,
which we call the Frobenius tensor, as follows: let $(E_{\ua})$ and
$(E^{\una})$ be bases for $\unF$ and $\unB^*$, the even cotangent bundle,
then the components of the Frobenius tensor in such a basis,  denoted by
$T_{\ua\ub}{}^{\unc}$, are given by

\beq T_{\ua\ub}{}^{\unc}=-\langle [E_{\ua}, E_{\ub}],E^{\unc}\rangle\ ,
\la{1ft} \eeq

where $\langle,\rangle$ denotes the usual pairing between vectors and
forms. This tensor is usually called the dimension zero torsion tensor but it
does not require the introduction of a connection to be well-defined. The
motivation for the nomenclature is that the vanishing of this tensor would imply
that $\unF$ is involutive so that the  Frobenius theorem would apply, i.e.
there would be purely odd integral submanifolds of $\unM$ whose tangent
spaces coincide with $\unF$. In the present context, this tensor is
actually maximally non-integrable in the sense that the even tangent bundle is
generated from the odd tangent bundle via the obvious map $\wedge^2
\unF\rightarrow \unB$ defined by the Frobenius tensor. From the definition
of the latter, this is related to taking commutators of odd vector fields
and so can be thought of as a statement  of supersymmetry.

If we assume that the above structure is irreducible, i.e. it does not break up
into the direct sum of two lower-dimensional ones, then there exists
a choice of basis such that the Frobenius tensor can be brought to the form

\beq T_{\ua\ub}{}^{\unc} =-i\left((\C^{\unc})_{\ua\ub}
+(\C^{\unb\unc})_{\ua\ub}X_{\unb\unc}{}^{\una}
+(\C^{\unb\unc\und\une\unf})_{\ua\ub}
Y_{\unb\unc\und\une\unf}{}^{\una}\right)\ . \la{1gf} \eeq

Here $X$ and $Y$ are both traceless and their totally antisymmetric parts
vanish when the ${\una}$ index is lowered using the standard
Lorentzian metric $\h_{\una\unb}$. Note that this is  a scale invariant
statement since it remains true for any metric in the same conformal class as
$\h$. This is the most general form of the Frobenius tensor possible in the
sense that it can be achieved by using all the freedom one has to choose the
bases $(E_{\ua}),\ (E^{\una})$ leaving only $CSpin(1,10):=Spin(1,10)\times
\bbR^*$ transformations.\footnote{In fact, one could have other signatures of
spacetime, but we shall not consider such cases here.} The supergravity
equations of motion are implied, at least locally, when the Frobenius
tensor takes the simple form

\beq T_{\ua\ub}{}^{\unc} =-i(\C^{\unc})_{\ua\ub}\ . \la{1t} \eeq

This is not immediately obvious but can be proven most simply by working
systematically through the Bianchi identities using the connection
formalism \cite{Howe:1997rf}.  One can restate this by saying that the
equations of motion of supergravity will follow if the Frobenius tensor is
invariant under $CSpin(1,10$. In practice, the tensors $X$ and $Y$ are non-zero
only when one includes higher derivative corrections to supergravity. In this
case they will be functions of the supergravity fields involving a dimensionful
parameter so that local scale invariance will be explicitly broken.

We now consider the embedding of the worldvolume $M$ in $\unM$. For the
2-brane $M$ is a $(3|16)$-dimensional supermanifold. The natural
superembedding condition, which holds for all branes at least at leading
order in derivatives, states that there is a choice of odd tangent bundle
$F$ for $M$ such that $F\subset \unF$ at all points of  $M$. Dually one
has  $B^*\subset \unB$. The only other requirement we need to impose on
the embedding is that the metric induced on $B^*$ (from $\h_{\una\unb}$)
should be Lorentzian. Again this condition is automatically conformally
invariant.

The embedding condition induces a Frobenius tensor on $M$ in
a straightforward manner. We introduce local bases $(E_{\a}),\ (E^a)$
for $F$ and $B^*$ respectively and note that $F\subset \unF$ implies that

\beq E_{\a}=E_{\a}{}^{\ua}E_{\ua} \la{1ealpha} \eeq

for some $16\xz 32$ matrix $E_{\a}{}^{\ua}$. Equivalently,

\be E_{\a}{}^{\una}=0\ . \la{1ec} \ee

Dually one has

\beq E^{\una}=E^a E_{a}{}^{\una} \la{1euna} \eeq

where on the left-hand side $E^{\una}$ is pulled-back onto $M$. One
then has

\bea \langle[E_{\a},E_{\b}],E^{\una}\rangle &=&
E_{\a}{}^{\ua}E_{\b}{}^{\ub}
\langle[E_{\ua},E_{\ub}],E^{\una}\rangle \nonumber \\
&=&\langle[E_{\a},E_{\b}],E^{a}\rangle E_a{}^{\una} \la{1eee} \eea

at any point $p\in M$. Thus the Frobenius tensors are related by

\beq E_{\a}{}^{\ua}E_{\b}{}^{\ub}T_{\ua\b}{}^{\unc}=T_{\a\b}{}^c
E_c{}^{\unc}\ . \la{frob} \eeq

If $(E_{\ua})$ is a spin basis for $\unF$ any other such basis will be
related to it by an element $u$ of $Spin(1,10)$ up to a conformal
factor which we shall ignore for the moment. We write
$u=(u_{\a}{}^{\ua},u_{\a'}{}^{\ua})$, with $\a'=1,\ldots 16$. Since
$E_{\a}{}^{\ua}$ has maximal rank there will be a choice of $u$ such
that $E_{\a}$ is related to $u_{\a}{}^{\ua}E_{\ua}$ by a non-singular
matrix. Hence, without loss of generality, we can write

\beq E_{\a}{}^{\ua}=A_{\a}{}^{\b}u_{\b}{}^{\ua}+
B_{\a}{}^{\b'}u_{\b'}{}^{\ua} \la{1ab} \eeq

where $\det A\neq 0$. Making a change of basis for $F$ we arrive at

\beq E_{\a}{}^{\ua}=u_{\a}{}^{\ua}+h_{\a}{}^{\b'}u_{\b'}{}^{\ua}\ .
\la{1h} \eeq

On the bosonic space $B^*$ the situation resembles more closely the
case of a Lorentzian embedding and we may choose, again up to a
conformal factor

\beq E_a{}^{\una}=u_a{}^{\una} \la{1u} \eeq

where $(u_a{}^{\una},u_{a'}{}^{\una})$ is the element of $SO(1,10)$
corresponding to $u=(u_{\a}{}^{\ua},u_{\a'}{}^{\ua})\in Spin(1,10)$.
Thus, at any point $p\in M$, the embedding is specified by
$u_a{}^{\una},u_{\a}{}^{\ua}$ and $h_{\a}{}^{\b'}$.

We can now decompose equation \eq{frob} into components tangent and
normal to $M$. We then find

\beq \unT_{\a\b}{}^{c'}+2h_{(\a}{}^{\c'}\unT_{\b)\c'}{}^{c'}+
h_{\a}{}^{\c'}h_{\b}{}^{\d'}\unT_{\c'\d'}{}^{c'}=0 \la{1trans} \eeq

and

\beq \unT_{\a\b}{}^{c}+2h_{(\a}{}^{\c'}\unT_{\b)\c'}{}^{c}+
h_{\a}{}^{\c'}h_{\b}{}^{\d'}\unT_{\c'\d'}{}^{c}=T_{\a\b}{}^c \la{1tan}
\eeq

where

\beq \unT_{\a\b}{}^{c'}=u_{\a}{}^{\ua}
u_{\b}{}^{\ub}T_{\ua\ub}{}^{\unc}u_{\unc}{}^{c'} \la{1indt} \eeq

and similarly for the other projections of $T_{\ua\ub}{}^{\unc}$. In
order for there to be embeddings of branes in general we require that
these equations be satisfied for arbitrary embeddings passing through a
given point $p\in \unM$ and furthermore that this should be true for
all points of $\unM$. Since we may vary $u$ and $h$ independently this
requires

\beq \unT_{\a\b}{}^{c'}=0\ . \la{1albec'} \eeq

This can only be satisfied for arbitrary embeddings if the
the target space tensor is invariant, i.e. if

\beq T_{\ua\ub}{}^{\unc}=-i(\C^{\unc})_{\ua\ub}\ . \la{1tgamma} \eeq

Given this one finds that equations \eq{1trans} and \eq{1tan} are
solved by (in the case of the supermembrane)

\beq h_{\a}{}^{\b'}=0 \la{1h0} \ee

while the worldvolume Frobenius tensor is

\be T_{\a\b}{}^c=-i(\C^c)_{\a\b}\ . \la{1wvf} \ee

As an aside we note that, in the case of the 5-brane, $h$ is no longer
zero. It is given by

\be h_{\a}{}^{\b'}= {1\over6}(\C^{abc})_{\a}{}^{\b'}h_{abc} \la{5brane}
\eeq

where $h_{abc}$ is self-dual. This field is related to the self-dual
3-form field strength of the worldvolume tensor multiplet. This a
general feature: $h_{\a}{}^{\b'}$ is non-vanishing whenever the
worldvolume multiplet includes a gauge field.

To summarise, in this subsection we have shown that the existence of
supermembranes as embedded submanifolds through arbitrary points of the
$(11|32)$-dimensional target superspace $\unM$ implies that the
equations of motion of eleven-dimensional supergravity must be
satisfied. Moreover,  the embedding condition determines the
worldvolume multiplet to be the on-shell $d=3,\ N=8$ scalar multiplet,
as we shall describe below. Thus the worldvolume and spacetime dynamics
are determined by the same simple embedding condition. This argument is
thus parallel to that given in the component formalism where it is
known that kappa-symmetry implies that the super target space must be a
solution of supergravity \cite{Bergshoeff:1987cm,Duff:1987bx}.


\subsection{Linearised embeddings}


In local coordinates, $z=(x,\th)$ for $M$ and
$\underline{z}=(\underline{x},\underline{\th})$ for $\unM$, the
embedding is given as $\underline{z}(z)$. The derivative of the
embedding, referred to preferred bases on both spaces, defines what we
shall call the embedding matrix, $E_A{}^{\unA}$,

\beq E_A{}^{\unA}=E_A{}^M\del_{M}z^{\unM}E_{\unM}{}^{\unA}\ , \la{1em}
\eeq

where $E_A{}^M$ is the inverse supervielbein on the worldvolume. In
order to get a feel for the components of the embedding matrix consider
the case of a flat target space,

\beqa E^{\una}&=&dx ^{\una} -{i\over2}
d\th^{\ua}(\C^{\una})_{\ua\ub}\th^{\ub}\ ,
\nonumber\\
E^{\ua}&=& d\th^{\ua}\ , \la{1fss} \eeqa

choose the physical gauge,

\beqa
x^{\una}&=&\cases{x^a &\cr x^{a'}(x,\th)  &\cr} \nonumber\\
&&\nn\\
\th^{\ua}&=&\cases{\th^{\a}&\cr\th^{\a'}&\cr} \la{1pg} \eeqa

and take the embedding to be infinitesimal so that $E_A{}^M\del_M$ can
be replaced by $D_A=(\del_a,D_{\a})$ where $D_{\a}$ is the flat
superspace covariant derivative on the worldvolume. In this limit the
embedding condition becomes

\beq D_{\a}X^{a'}=i(\C^{a'})_{\a\b'}\th^{\b'} \la{1lec} \eeq

where

\beq X^{a'}:=x^{a'}+{i\over2}\th^{\a}(\C^{a'})_{\a\b'}\th^{\b'}\ .
\la{1xtrans} \eeq

One also finds

\beqa E_a{}^{\unb}&\rightarrow&\cases{\d_a{}^b &\cr\del_a
X^{b'}&\cr}\nonumber\w2 E_a{}^{\ub}&\rightarrow&\cases{0 &\cr
\del_a\th^{\b'} &\cr}\nonumber\w2
E_{\a}{}^{\ub}&\rightarrow&\cases{\d_{\a}{}^{\b}&\cr D_{\a}\th^{\b'} &
\cr} \la{1ecomp} \eeqa

One can therefore think of the leading components of $E_a{}^{\una}$ and
$E_a{}^{\ua}$ as being the spacetime derivatives of the world surface
scalar and spinor fields (i.e. $X^{a'}$ and $\th^{\a'}$ in the
linearised case). From equation \eq{1lec} we find, by differentiating,
that

\beq D_{\a}\th^{\b'}={1\over2}(\C^{ab'})_{\a}{}^{\b'}\del_a X_{b'}\ .
\la{1dth} \eeq

This shows that the only independent components of the Goldstone
superfield $X^{a'}$ are the eight scalars given by $X^{a'}|$ and
$\th^{\a'}|$ where the vertical bar denotes  the evaluation of a
superfield at $\th=0$. The worldvolume multiplet of the linearised
superembedding is therefore the free $d=3,N=8$ on-shell scalar
multiplet.

It is instructive to consider this type of embedding to next order in
the transverse coordinates $(x^{a'},\th^{\a'})$. We can write

\beq E_{A}=D_A - H_{A}{}^B D_B \la{1H} \eeq

where $H_A{}^B$ is the linearised supervielbein. To first order in the
transverse coordinates it is easy to show that $H$ vanishes (in the
physical gauge), but at second order one finds, from \eq{1ec}

\beq H_{\a}{}^b=-{i\over2}D_{\a}\th^{\b'}(\C^b)_{\b'\c'}\th^{\c'}\ .
\eeq

Since one can determine the remaining components of $H_A{}^B$ from the
constraints on the worldvolume torsion, this shows that the induced
geometry of the worldvolume is indeed determined by the embedding as
claimed.


\section{Lorentzian supergeometry and superembeddings}


\subsection{Target superspace geometry}


The discussion of the previous section makes no reference to
connections, torsion or curvature, but in order to interpret the
equations given there it is convenient to introduce these objects in
order to obtain manifestly covariant equations of motion for the
component fields. Our approach is similar to that of
\cite{Bandos:1995zw}, although we work in an arbitrary supergravity
background and give more details of the equations of motion and the
induced supergeometry on the brane. We begin with the target space
$\unM$. We first make a choice of the even tangent bundle $\unB$ (i.e.
fix $(E_{\una})$) which reduces the structure group to
$Spin(1,10)\xz\bbR^{+}:=CSpin(1,10)$. We then introduce a connection,
$\underline{\O}$, but only for $Spin(1,10)$, since the scale part of
the group can be dealt with separately. We define the torsion and
curvature 2-forms as usual by

\beqa T^{\unA}          &=&   d E^{\unA} + E^{\unB}
\O_{\unB}{}^{\unA}\nonumber\w2 R_{\unA}{}^{\unB} &=&
d\O_{\unA}{}^{\unB}+\O_{\unA}{}^{\unC}\O_{\unC}{}^{\unB}\ , \la{3TR}
\eeqa

where the basis forms $(E^{\unA})$ are related to the coordinate basis
by the vielbein matrix, $E_{\unM}{}^{\unA}$,

\beq E^{\unA}=dz^{\unM} E_{\unM}{}^{\unA} \la{3vm} \eeq

whose inverse is denoted $E_{\unA}{}^{\unM}$.

The dimension zero component of the torsion tensor is identified with
the Frobenius tensor of the last section and is thus required to be, if
embeddings of branes are to be permitted,

\beq T_{\ua\ub}{}^{\unc}=-i(\C^{\unc})_{\ua\ub}\ . \la{3t} \eeq

The dimension one-half component of the connection and $E_{\una}$ can
be chosen such that almost all components of the dimension one-half
torsion vanish except for a single spinor field. However, it can be
shown that this spinor is the odd derivative of a scalar superfield, at
least provided that $\unM$ is simply connected, and that this scalar
field can be transformed away by a super-Weyl transformation. We shall
assume throughout the rest of this paper that this manoeuvre is both
allowable and has been carried out  so that

\beq T_{\ua\unb}{}^{\unc}=T_{\ua\ub}{}^{\uc}=0\ . \la{3dht} \eeq

At dimension one one may impose

\beq T_{\una\unb}{}^{\unc}=0 \la{3tabc} \eeq

in which case one finds

\beq T_{\una\ub}{}^{\uc}=-
{1\over36}(\C^{\unb\unc\und})_{\ub}{}^{\uc}G_{\una\unb\unc\und}
-{1\over288}(\C_{\una\unb\unc\und\une})_{\ub}{}^{\uc}G^{\unb\unc\und\une}
\ . \la{3td1} \eeq

The dimension three-halves torsion, whose leading component in the
$\underline{\theta}$-expansion is the field strength tensor for the
gravitino field, is given as the odd derivative of the field $G$. The
curvature can be computed in terms of the torsion, i.e. in terms of $G$
and its derivatives, by virtue of Dragon's theorem which is applicable
in this case. Finally, given the above, one can establish the existence
of a closed superspace four-form $G_4$ whose non-vanishing components
are $G_{\una\unb\unc\und}$ and

\beq G_{\una\unb\uc\ud}=-i(\C_{\una\unb})_{\uc\ud}\ . \la{3g4} \eeq


\subsection{Superembeddings}


The bosonic degrees of freedom are all scalar. In the context of
superembeddings this implies the vanishing of the field
$h_{\a}{}^{\b'}$ as we saw previously, so that

\be E_{\a}{}^{\ua}=u_{\a}{}^{\ua}\ ,\quad \quad
T_{\a\b}{}^c=-i(\C^c)_{\a\b}\ . \la{3et} \ee

The first task is to specify the various bundles that arise. This is
more complicated than the Riemannian case where one has only to fix the
normal bundle which can be done by choosing it to be the orthogonal
complement of the tangent bundle of the surface. In the supersymmetric
case we have the additional complication of even and odd bundles, so
that there are nine bundles in all to think about; odd, even and total
for the target space, the worldvolume and the normal direction to the
worldvolume. We shall assume that the geometry of the target space is
standard so that a decomposition $\unT=\unF\oplus\unB$ for the target
space bundles has been chosen such that equations \eq{3dht} hold. It
therefore remains to specify how $\unT$  decomposes in terms of
tangential and normal components.

It is natural to take the odd normal bundle $F'$ to be a subbundle of
$\unF$ complementary  to $F$. Dually, one can take $B'^{*}$ to be a
subbundle of $\unB^*$ complementary  to $B^*$. We introduce a basis
$(E_{A'})=(E_{a'},E_{\a'})$ for the normal bundle $T'$  related to the
preferred basis for the target space by the normal matrix,
$E_{A'}{}^{\unA}$,

\beq E_{A'}=E_{A'}{}^{\unA} E_{\unA}\ , \la{3nm} \eeq

The assumption that $F'$ is a subspace of $\unF$ is equivalent to

\beq E_{\a'}{}^{\una}=0\ . \la{3nec} \eeq

$F'$ and $B'^*$ can be specified completely using the Frobenius tensor.
We may choose

\beqa
E_{\a}{}^{\ua}&=&u_{\a}{}^{\ua}\nno\\
E_{\a'}{}^{\ua}&=&u_{\a'}{}^{\ua} \la{3ff} \eeqa

and

\beqa
E_{a}{}^{\una}&=&u_{a}{}^{\una}\nno\\
E_{a'}{}^{\una}&=&u_{a'}{}^{\una} \la{3bb} \eeqa

where $(u_{\a}^{\ua},u_{\a'}{}^{\ua})$ is an element of $Spin(1,10)$
and $(u_a{}^{\una},u_{a'}{}^{\una})$ is the corresponding element of
the Lorentz group. This is accomplished by demanding that the Frobenius
tensor decompose as follows

\beq T_{\ua\ub}{}^{\unc}\rightarrow\cases{ T_{\a\b}{}^{c}
&$=-i(\C^{c})_{\a\b}$\cr T_{\a'\b'}{}^{c} &$=-i(\C^{c})_{\a'\b'}$\cr
T_{\a\b'}{}^{c'} &$=-i(\C^{c'})_{\a\b'}=T_{\b'\a}{}^{c'}$\cr}
\la{3frob} \eeq

with all other components of $T_{\ua\ub}{}^{\unc}$ being zero. Since we
have a Lorentzian metric on $\unB$, the conformal factor allowed by
considerations of the Frobenius tensor alone can be set equal to one.
This is implicit in the expressions for $E_a{}^{\una}$ and
$E_{a'}{}^{\una}$ given above. Equivalently we could impose

\beq E_a{}^{\una} E_b{}^{\unb}\h_{\una\unb}=\h_{ab} \la{3cf} \eeq

and

\beq E_{a'}{}^{\una} E_{b'}{}^{\unb}\h_{\una\unb}=\d_{a'b'}\ .
\la{3cfn} \eeq

The embedding matrix together with the normal matrix makes up a square
matrix with entries $(E_A{}^{\unA},E_{A'}{}^{\unA})$ whose inverse will
be denoted by $(\iE_{\unA}{}^{A},\iE_{\unA}{}^{A'})$. In view of the
embedding condition \eq{1ec} one finds that
$(\iE_{\ua}{}^{\a},\iE_{\ua}{}^{\a'})$ is the inverse of
$(E_{\a}{}^{\ua},E_{\a'}{}^{\ua})$ and that
$(\iE_{\una}{}^{a},\iE_{\una}{}^{a'})$ is the inverse of
$(E_{a}{}^{\una},E_{a'}{}^{\una})$. Furthermore,

\beq \iE_{\ua}{}^a=\iE_{\ua}{}^{a'}=0\ . \la{3ee0} \eeq

The inverse is completed by

\beqa \iE_{\una}{}^{\a}&=& \iE_{\una}{}^a E_a{}^{\ua} \iE_{\ua}{}^{\a}
+ \iE_{\una}{}^{a'} E_{a'}{}^{\ua} \iE_{\ua}{}^{\a}\nonumber\w2
\iE_{\una}{}^{\a'}&=& \iE_{\una}{}^a E_a{}^{\ua} \iE_{\ua}{}^{\a'} +
\iE_{\una}{}^{a'} E_{a'}{}^{\ua} \iE_{\ua}{}^{\a'}\ . \la{3einv} \eeqa

The embedding condition does not fully specify the even tangent bundle
$B$ of $M$. Making a choice of $B$ is equivalent to choosing a basis
set of even vectors, $(E_a)$, which can be modified by

\beq E_a\mapsto E_a +\r_a{}^{\a} E_{\a}\ . \la{3rho} \eeq

>From an embedding point of view one has

\beq E_a=E_a{}^{\una} E_{\una}+E_a{}^{\ua} E_{\ua}\ . \la{3ea} \eeq

The second term can be written as

\beq E_a{}^{\ua}E_{\ua}= E_{a}{}^{\ua} \iE_{\ua}{}^{\a}E_{\a} +
E_{a}{}^{\ua} \iE_{\ua}{}^{\a'}E_{\a'}\ . \la{3ee2} \eeq

By using a redefinition of the form \eq{3rho}, one can clearly choose
$\r$ so that the first term on the right in \eq{3ee2} is removed. Thus
we can always make the standard choice of $B$ for which

\beq E_a{}^{\ua}=\L_a{}^{\a'} E_{\a'}{}^{\ua}\ . \la{3L} \eeq

Finally we can choose $B'$ to be the orthogonal complement of
$B\cap\unB$ and to have vanishing projection on $\unF$.

In summary, we may choose the basis vectors for the  spaces $F,B,F'$
and $B'$ by specifying the components of the embedding and normal
matrices to be

\beq \barr{lcllcl} E_a{}^{\una}&=&u_{a}{}^{\una}\qquad& E_a{}^{\ua}&=&
\L_a{}^{\a'}
u_{\a'}{}^{\ua}\\
E_{\a}{}^{\una}&=& 0 & E_{\a}{}^{\ua}&=& u_{\a}{}^{\ua} \earr \la{3em}
\eeq

and

\beq \barr{lcllcl}
E_{a'}{}^{\una}&=&  u_{a'}{}^{\una}\qquad& E_{a'}{}^{\ua}&=&  0\\
E_{\a'}{}^{\una}&=& 0 & E_{\a'}{}^{\ua}&=& u_{\a'}{}^{\ua} \earr
\la{3em2} \eeq

respectively. One can then compute the components of the inverse
matrices to be

\beq \barr{lcllcl} \iE_{\una}{}^{a}&=&
u_{\una}{}^{a}\qquad&\iE_{\una}{}^{\a\phantom{'}}&=& 0
\\
\iE_{\ua}{}^{a}&=& 0 & \iE_{\ua}{}^{\a}&=&
u_{\ua}{}^{\a}\phantom{\L_a{}^{\a'}} \earr \la{3einvm} \eeq

and

\beq \barr{lcllcl} \iE_{\una}{}^{a'}&=&  u_{\una}{}^{a'}\qquad&
\iE_{\una}{}^{\a'}
&=& -u_{\una}{}^a \L_a{}^{\a'}\\
\iE_{\ua}{}^{\a'}&=& u_{\ua}{}^{\a'} & \iE_{\ua}{}^{a'}&=& 0\ . \earr
\la{3einvm2} \eeq

The structure groups induced on the worldvolume bundles by the
embedding are $Spin(1,2)\cdot Spin(8)$ on $F$ and $F'$ (but acting via
the two inequivalent minimal representations of $Spin(8)$), $SO_o(1,2)$
on $B$ (with the superscript ``o'' denoting the component connected to
the identity) and $SO(8)$ on $B'$, the orthogonal groups being related
to the spin groups by the $2:1$ maps determined by the $\C$-matrices.

We shall now turn to the induced connection on the worldvolume. This
can be specified by means of the superspace version of the
Gauss-Weingarten equations which were discussed in
\cite{Bandos:1995zw}. These are summarised in the Riemannian case in
\eq{gw} in the appendix. The super Gauss-Weingarten equations are

\beqa
\nab_A E_b{}^{\unc} &=& K_{A,b}{}^{c'} E_{c'}{}^{\unc}\nno\\
\nab_A E_{b'}{}^{\unc} &=& K_{A,b'}{}^{c} E_{c}{}^{\unc} \la{3gw1}
\eeqa

and

\beqa
\nab_A E_{\b}{}^{\uc} &=& K_{A,\b}{}^{\c'} E_{\c'}{}^{\uc}\nno\\
\nab_A E_{\b'}{}^{\uc} &=& K_{A,\b'}{}^{\c} E_{\c}{}^{\uc} \la{3gw2}
\eeqa

where

\beqa \nab_A E_{b}{}^{\unc}&=& E_A
E_{b}{}^{\unc}-\O_{A,b}{}^{c}E_{c}{}^{\unc} +
E_b{}^{\unb}\O_{A,\unb}{}^{\unc}\nn \w2 \nab_A E_{b'}{}^{\unc}&=& E_A
E_{b'}{}^{\unc}-\O_{A,b'}{}^{c'}E_{c'}{}^{\unc} +
E_{b'}{}^{\unb}\O_{A,\unb}{}^{\unc} \nn\w2 \nab_A E_{\b}{}^{\uc}&=& E_A
E_{\b}{}^{\uc}-\O_{A,\b}{}^{\c}E_{\c}{}^{\uc} +
E_{\b}{}^{\ub}\O_{A,\ub}{}^{\uc} \nn\w2 \nab_A E_{\b'}{}^{\uc}&=& E_A
E_{\b'}{}^{\unc}-\O_{A,\b'}{}^{\c'}E_{\c'}{}^{\uc} +
E_{\b'}{}^{\ub}\O_{A,\ub}{}^{\uc}\ . \la{3def} \eeqa

In these equations $\O_{A,b}{}^{c}$, etc. are connections on the
worldvolume while $\O_{A,\unb}{}^{\unc}$ and $\O_{A,\ub}{}^{\uc}$ are
the pull-backs of the target space connections in $\unB$ and $\unF$.
For example,

\beq \O_{A,\unb}{}^{\unc}=E_A{}^{\unA}\O_{\unA,\unb}{}^{\unc}\ .
\la{3eg} \eeq

Note that the Gauss-Weingarten equations as defined above differ
slightly from the Riemannian case in that not all components of the
embedding matrix appear. Indeed, if $Y$ is an even vector on $M$, $Y^a
E_a{}^{\una}$ does not give the components of $Y$ considered as a
vector on $\unM$, but only the projection of this vector onto $\unB$.

Because of the $Spin(1,10)$ structure on the target space we have

\beq
\O_{A,\ub}{}^{\uc}=\frac{1}{4}(\C^{\unb\unc})_{\ub}{}^{\uc}\O_{A,\unb\unc}\
. \la{3off} \eeq

It is not difficult to show that the induced connections preserve the
various tensors that are induced on the worldvolume bundles from the
metrics and the Frobenius tensor on $M$, so that one has the following
relations between the connections

\beqa
\O_{A,bc}&=&-\O_{A,cb}\nn\\
\O_{A,b'c'}&=&-\O_{A,c'b'}\nn\\
\O_{A,\b}{}^{\c}&=&\frac{1}{4}(\C^{bc})_{\b}{}^{\c}\O_{A,bc}
+\frac{1}{4}(\C^{b'c'})_{\b}{}^{\c}\O_{A,b'c'}\nn\\
\O_{A,\b'}{}^{\c'}&=&\frac{1}{4}(\C^{bc})_{\b'}{}^{\c'}\O_{A,bc}
+\frac{1}{4}(\C^{b'c'})_{\b'}{}^{\c'}\O_{A,b'c'} \la{3conn} \eeqa

while the components of the second fundamental form obey

\beqa
K_{A b,c'}&=&-K_{A,c',b}\nn\\
K_{A,\b}{}^{\c'}&=&\frac{1}{2}(\C^{bc'})_{\b}{}^{\c'}K_{A,bc'}\nn\\
K_{A,\b'}{}^{\c}&=&\frac{1}{2}(\C^{bc'})_{\b'}{}^{\c}K_{A,bc'}\ .
\la{3K} \eeqa

The connections can equally well be defined from the
$\gs\gp\gi\gn(1,10)$-valued one-form

\beq K_A=(\nab_A u)u^{-1} \la{3Kform} \eeq

where $u=(u_{\a}{}^{\ua},u_{\a'}{}^{\ua})$, and where $\nab_A$ acts on
$u$ as in \eq{3def}. The 1-form $K$ can be defined for any worldvolume
connection; the induced connection is given by demanding that the
$\gs\gp\gi\gn(1,2)$ and $\gs\gp\gi\gn(8)$ components vanish. The
remaining non-zero components then satisfy equations \eq{3K}.

The equation defining the target space torsion 2-form, pulled back to
the worldvolume, reads

\beq \nab_{A} E_{B}{}^{\unC}-(-1)^{AB} \nab_{B} E_{A}{}^{\unC}
+T_{AB}{}^C E_C{}^{\unC}= (-1)^{(A(B+\unB)}
 E_B{}^{\unB} E_A{}^{\unA} T_{\unA\unB}{}^{\unC}\ .
\la{3tor} \eeq

Finally, by differentiating \eq{3Kform}, we obtain

\beq \nab_A K_B-\nab_B K_A +T_{AB}{}^C K_C-[K_A,\,K_B]=(R_{AB}u)u^{-1}
\la{3dK} \eeq

where the curvature form acts on the indices of $u$ according to the
prescription given in \eq{3def}. For the case of induced connections,
one finds the equations of Gauss and Codazzi,

\beqa R_{AB,c}{}^{d} &=& \unR_{AB,c}{}^{d} +
(K_{A,c}{}^{e'}K_{B,e'}{}^d\mp (A\leftrightarrow B))\nonumber \w2
R_{AB,c'}{}^{d'} &=& \unR_{AB,c'}{}^{d'} +
(K_{A,c,}{}^{e}K_{B,e}{}^{d'}\mp (A\leftrightarrow B)) \la{3gc} \eeqa

where

\beq \unR_{AB,c}{}^{d}=(-1)^{A(B+\unB)}E_B{}^{\unB} E_A{}^{\unA}
u_c{}^{\unc}\,R_{\unA\unB,\unc}{}^{\und}\, u_{\und}{}^{d} \la{3R1} \eeq

and, similarly,

\beq \unR_{AB,c'}{}^{d'}=(-1)^{A(B+\unB)}E_B{}^{\unB} E_A{}^{\unA}
u_{c'}{}^{\unc}\,R_{\unA\unB,\unc}{}^{\und}\, u_{\und}{}^{d'}\ .
\la{3R2} \eeq

The Gauss-Codazzi equations given above are enough to define  all of
the curvature components because those with spinor indices are
determined by those with vector indices according to the relations
\eq{3conn} which hold also for the curvature forms.

The  torsion equation \eq{3tor} can be used to determine the induced
worldvolume torsion as well as the constraints satisfied by the
superfield $\L$. The curvature components can be read off using the
Gauss-Codazzi equation.


\section{The dynamics of the M2-brane}


\subsection{The torsion equations}


In this section we analyse the membrane embedding by systematically
going through the torsion equation \eq{3tor}. We shall take the
worldvolume connection to be induced and the embedding to be of the
form given in equations \eq{3em} and \eq{3em2}. The analysis will
determine the worldvolume torsion and the equations of motion of the
membrane in an arbitrary supergravity background. The worldvolume
curvature can be found using the Gauss-Codazzi equations.

In natural units $E_{\a}{}^{\ua}$ and $E_a{}^{\una}$ have dimension
zero, $E_{\a}{}^{\una}$ has dimension minus one-half and $E_a{}^{\ua}$
has dimension plus one-half. The dimension zero component of the
torsion equation simply reduces to the relation \eq{frob} between the
Frobenius tensors of the worldvolume and the target space, and this has
already been taken account of in writing down the form of the embedding
matrix \eq{3em}.

At dimension one-half one has two equations,

\beq \nab_{\a} E_{\b}{}^{\uc}+\nab_{\b} E_{\a}{}^{\uc} +T_{\a\b}{}^c
E_c{}^{\uc} +T_{\a\b}{}^{\c}E_{\c}{}^{\uc}=0 \la{4dht1} \eeq

and

\beq \nab_{\a} E_b{}^{\unc} +T_{\a b}{}^c E_c{}^{\unc}=-E_b{}^{\ub}
E_{\a}{}^{\ua} T_{\ua\ub}{}^{\unc}=iE_b{}^{\ub} E_{\a}{}^{\ua}
(\C^{\unc})_{\ua\ub} \la{4dht2} \eeq

where the vanishing of $E_{\a}{}^{\una}$ has been taken into account.
These can be analysed by right-multiplying  by $u^{-1}$ in the
appropriate representation. Right-multiplying \eq{4dht1} by
$u_{\uc}{}^{\c}$ one finds

\beq T_{\a\b}{}^{\c}=0 \la{4wvt1} \eeq

where we have used the induced connection condition
$K_{\a,\b}{}^{\c}=0$ and the embedding choice \eq{3L}. Similarly,
right-multiplying \eq{4dht2} by $u_{\unc}{}^c$, one finds

\beq T_{\a b}{}^c=0 \la{4wvt2} \eeq

Right-multiplying \eq{4dht2} with $u_{\unc}{}^{c'}$ gives

\beq K_{\a,bc'}=(\C_{c'})_{\a\b'}\L_b{}^{\b'} \la{4K1} \eeq

from which one can deduce, using \eq{3K}, that

\beq K_{\a,\b}{}^{\c'}=
{i\over2}(\C^{ab'})_{\b}{}^{\c'}(\C_{b'})_{\a\d'}\L_a{}^{\d'}\ .
\la{4K2} \eeq

The remaining dimension one-half equation is found by right-multiplying
\eq{4dht1} by $u_{\uc}{}^{\c'}$. This gives

\beq K_{\a,\b}{}^{\c'} + K_{\b,\a}{}^{\c'}=
i(\C^c)_{\a\b}\L_{c}{}^{\c'}\ . \la{4K3} \eeq

Due to \eq{4K2} this relation involves only the field $\L_a{}^{\a'}$;
it is satisfied provided that

\beq (\C^a)_{\a\b'}\L_a{}^{\b'}=0 \ . \la{4de} \eeq

In view of the linearised analysis, one sees that the leading component
of this equation is the Dirac equation for the worldvolume fermion
field.

There are also two equations at dimension one:

\beq \nab_a E_{\b}{}^{\uc}-\nab_{\b} E_a{}^{\uc} +T_{a\b}{}^{\c}
E_{\c}{}^{\uc}=E_{\b}{}^{\ub} E_a{}^{\una} T_{\una\ub}{}^{\unc}=0
\la{4d1t1} \eeq

and

\beq \nab_a E_b{}^{\unc}-\nab_b E_a{}^{\unc} +T_{ab}^c
E_c{}^{\unc}=E_b{}^{\ub} E_a{}^{\ua} T_{\ua\ub}{}^{\una}\ . \la{4d1t2}
\eeq

Right-multiplying by $u$ as above one finds, from \eq{4d1t1},

\beqa T_{a\b}{}^{\c}&=&\unT_{a\b}{}^{\c}-\L_a{}^{\d'} K_{\b,\d'}{}^{\c}
\la{4wvt3}\w2 K_{a,\b}{}^{\c'}-\nab_{\b} \L_a{}^{\c'}&
=&\unT_{a\b}{}^{\c'}\ . \la{4K4} \eeqa

>From \eq{4d1t2} one finds

\beqa T_{ab}{}^c&=&i\L_a{}^{\a'}(\C^c)_{\a'\b'}\L_b{}^{\b'}\la{4wvt4}
\w2 K_{a,b}{}^{ c'}&=&K_{b,a}{}^{c'}\ . \la{4K5} \eeqa

The first and third of these equations determine the dimension one
components of the worldvolume torsion while the fourth shows that the
bosonic part of the second fundamental form is symmetric even though
$T_{ab}{}^c$ is not zero. The main equation to analyse is therefore the
second one. To do this we shall switch to the two step notation (see
appendix) in which worldvolume spinor indices are explicitly decomposed
into representations of $SL(2,\bbR)\xz O(8)$. We may set

\beq \nab_{\b}\L_a{}^{\c'}\rightarrow\nab_{\b j}\L_a{}^{\c
k'}=\d_{\b}{}^{\c}M_{a j}{}^{k'} +(\c^b)_{\b}{}^{\c}N_{abj}{}^{k'} \eeq

The Dirac equation for $\L$ implies

\beqa
\h^{ab}N_{ab jk'}&=& 0\nonumber\\
M_{ajk'}&=&\e_{abc} N^{bc}_{jk'}\ . \la{4mn} \eeqa

Substituting for $\nab\L$ in terms of $M$ and $N$ in \eq{4K4} one finds
the following further conditions

\beqa N_{(ab)jk'}&=&
{1\over2}(\s^{c'})_{jk'}(K_{ab,c'}-{1\over3}\h_{ab}\h^{cd}K_{cd,c'})\nn\w2
M_{a,jk'}&=&+{1\over36}(\s^{b'c'd'})_{jk'}\unG_{ab'c'd'} \la{4mn2}
\eeqa

as well as

\beq \h^{ab} K_{ab,d'}=-{1\over6}\e^{abc}\unG_{abc}{}^{d'} \ .
\la{4kge} \eeq

Since $K_{a,b}{}^{c'}$ is essentially the second derivative of the
worldvolume scalar field, these equations show that all dimension one
quantities are determined in terms of known fields and their
derivatives. Equation \eq{4kge}, or rather its leading component, is
the scalar equation of motion. Using the fact that $K_{a,bc}=0$, the
definition of $K$ and noting that $\unG_{abcd}=0$ identically, one can
rewrite it in the form

\beq \h^{ab}\nab_a
E_{b}{}^{\und}=-{1\over6}\e^{abc}\unG_{abc}{}^{\und}\ . \la{4kge2} \eeq

This is the scalar equation in compact form in the presence of a
supergravity background. We remind the reader that, throughout this
section, we have used the notation that underlined tensors with
non-underlined indices are obtained from the target space tensors by
projecting with suitable factors of the $u$ matrices.

There is one equation at dimension three-halves which can be used to
determine the dimension three-halves worldvolume torsion component.
Finally, one can use the Gauss-Codazzi equations \eq{3gc} to determine
the worldvolume curvature directly.


\subsection{The equations of motion in component form}


In order to derive the component form of the equations of motion it is
necessary to relate the components of the worldvolume supervielbein to
known quantities, at least to leading order in $\th$. It is always
possible to choose a gauge in which the supervielbein takes the form

\beq \barr{lcllcl} E_m{}^a(x,\th)| &=&e_m{}^a(x)\qquad &
E_m{}^{\a}(x,\th)|&=&\psi_m{}^{\a}(x)\w2 E_{\m}{}^a(x,\th)|&=&0\qquad &
E_{\m}{}^{\a}(x,\th)|&=&\d_{\m}{}^{\a}\ , \earr \la{4sv} \eeq

where the vertical bar indicates evaluation at $\th=0$. The inverse
supervielbein to leading order is given by

\beq \barr{lcllcl} E_a{}^m(x,\th)| &=&e_a{}^m(x)\qquad &
E_a{}^{\m}(x,\th)|&=&-\psi_a{}^{\m}(x)\w2 E_{\a}{}^m(x,\th)|&=&0\qquad&
E_{\a}{}^{\m}(x,\th)|&=&\d_{\a}{}^{\m}\ ,\earr \la{4invsv} \eeq

where

\beq \psi_a{}^{\m}=e_a{}^m\psi_m{}^{\a}\d_{\a}{}^{\m} \ . \la{4psi}
\eeq

For the connection we can choose

\beq \O_m|=\o_m(x), \qquad\O_{\m}|=0 \la{4conn} \eeq

so that

\beq \O_{a}|=\o_a:=e_a{}^m \o_m, \qquad\O_{\a}|=0\ . \la{4conn2} \eeq

It is not difficult to see that the worldvolume vielbein $e_m{}^a$ is
the vielbein associated with the induced Green-Schwarz (GS) metric,

\beq g_{mn}={\cE}_m{}^{\una}{\cE}_n{}^{\unb}\h_{\una\unb} \la{4gmn}
\eeq

where

\beq {\cE}_m{}^{\una}:=\del_m z^{\unM} E_{\unM}{}^{\una}| \ .
\la{4gsem} \eeq

This is because $E_{\a}{}^{\una}=0$ implies, (from the definition
\eq{1em}) that

\beq \del_{\m}z^{\unM} E_{\unM}{}^{\una}|=0 \la{4dz} \eeq

and so

\beq E_a{}^{\una}|=e_a{}^m {\cE}_m{}^{\una}\ . \la{4ebb} \eeq

Since $E_a{}^{\una}$ is orthonormal with respect to the target space
metric it follows that $e_m{}^a$ is indeed the vielbein for the GS
metric.

For the worldvolume gravitino we have

\beqa \psi_m{}^{\a}&=&\del_m z^{\unM} E_{\unM}{}^{\unA}
E_{\unA}{}^{\a}|\nn\w2 &=&{\cal E}_m{}^{\ua} v_{\ua}{}^{\a} \la{4psi2}
\eeqa

where $v=u|$ (for any relevant pair of indices), and where

\beq {\cE}_m{}^{\ua}:=\del_m z^{\unM} E_{\unM}{}^{\ua}|\ . \la{4dth}
\eeq

This follows from the fact that $E_{\una}{}^{\a}=0$.

For the embedding matrix, we find, at leading order,

\beq \barr{lcllcl} E_{a}{}^{\ua}|&=&{\cE}_a{}^{\ub}
Q_{\ub}{}^{\ua}\qquad &
E_{a}{}^{\una}|&=&{\cE}_a{}^{\una}=v_a{}^{\una}\w2 E_{\a}{}^{\una}|
&=&0\qquad& E_{\a}{}^{\ua}|&=&v_{\a}{}^{\ua} \earr \la{4gsem2} \eeq

where

 \beqa
{\cE}_{a}{}^{\una}&=&e_a{}^m{\cE}_m{}^{\una}\nn\w2
{\cE}_{a}{}^{\ua}&=&e_a{}^m{\cE}_m{}^{\ua} \la{4defs} \eeqa

and where the projection operator $Q$ is given by

\beq Q_{\ua}{}^{\ub}=v_{\ua}{}^{\c'} v_{\c'}{}^{\ub}:=\frac{1}{2}(1-\C)
\ . \la{4Q}\eeq

We shall denote the complementary projector by $P$,

\beq P_{\ua}{}^{\ub}=v_{\ua}{}^{\c} v_{\c}{}^{\ub}= \frac{1}{2}(1+\C)
\la{4P} \eeq

We next consider the spin one-half equation of motion which is the
leading component of \eq{4de}. Consider

\beq E_a{}^{\ua}(\C^a)_{\ua\ub}=\L_a{}^{\a'}(\C^a)_{\a'\ub}=0 \la{4de2}
\eeq

where the last step follows by the equation of motion if $\ub=\b'$ and
by the properties of the gamma-matrices if $\ub=\b$. Evaluating this at
$\th=0$ and using \eq{4gsem2} one finds the component spin-half
equation

\beq {\cE}_a{}^{\ub}Q_{\ub}{}^{\ua}(\C^a)_{\ua\uc}=0 \la{4de3} \eeq

or, equivalently,

\beq g^{mn}{\cE}_m{}^{\uc}(1-\C)_{\uc}{}^{\ub}(\C_m)_{\ub\ua}=0
\la{4de4} \eeq

where $g$ is the induced Green-Schwarz metric and $\C_m=e_m{}^a\C_a$.

It is necessary to work a little harder to get the scalar equation of
motion out in component form. It is given by the leading component of
the superspace equation \eq{4kge2}. One has

\beq \nab_a E_b{}^{\und}=E_a{}^m\del_m
E_b{}^{\und}+E_a{}^{\m}\del_{\m}E_b{}^{\und}+\ldots \la{4dele} \eeq

where the dots denote connection terms. To evaluate  the second term on
the right-hand side we note that

\beq \nab_{\a}E_b{}^{\und}=E_{\a}{}^{\m}\del_{\m}
E_{\b}{}^{\und}+\ldots \la{4dele2} \eeq

where the additional terms vanish at leading order in $\th$. Since

\beqa \nab_{\a} E_b{}^{\unc}&=&K_{\a b}{}^{\c'} u_{c'}{}^{\unc}\nn\w2
   &=& (\C^{c'})_{\a\b'}\L^{\b'}u_{c'}{}^{\unc}\ ,
\la{4dele3} \eeqa

this term can be evaluated straightforwardly. One finds

\beq \h^{ab}E_a{}^{\m}\del_{\m}
E_b{}^{\und}|=-i\h^{ab}\psi_a{}^{\a}(\C^{c'})_{\a\b'}\l_b{}^{\b'}
v_{c'}{}^{\und} \la{4dele4} \eeq

where $\l=\L|$.  The connection terms in \eq{4dele2} involving the
fermionic part of the connection $\O_{\m}$ vanish at leading order and
are therefore irrelevant. There are two bosonic connection terms, one
for the (induced) worldvolume connection and one for the pulled-back
target space connection. The latter is easily taken into account since
it simply acts on underlined indices at $\th=0$ via the Green-Schwarz
pull-back, that is, the pull-back from the target superspace to the
bosonic worldvolume. It is the former which needs careful evaluation.
The term in question is $\h^{ab}(-E_a{}^m\O_{m,b}{}^{\und})|$. Now

\beq T_{ab}{}^c=(-E_a{}^M\del_M E_b{}^N E_N{}^c +\O_{a,b}{}^c)  -
(a\leftrightarrow b) \la{4tabc} \eeq

so that, at $\th=0$, we have

\beq T_{ab}{}^c|=\left(-e_a{}^m\del_m e_b{}^n e_n{}^c
-(E_a{}^{\m}\del_{\m}E_b{}^n)|e_n{}^c +\o_{a,b}{}^c\right) -
(a\leftrightarrow b)\ . \la{4tabc2} \eeq

This equation can be solved straightforwardly for $\o$; the terms
involving the vielbein alone give the Levi-Civita connection in an
orthonormal basis associated with the GS metric, while the remaining
terms give bilinear fermion contributions. The dimension one torsion is
\eq{4wvt4}

\beq T_{ab}{}^c=i\L_a{}^{\a'}(\C^c)_{\a'\b'}\L_b{}^{\b'} \la{4tabc3}
\eeq

so that the left-hand side of \eq{4tabc2} is easily evaluated. We also
need to evaluate $E_a{}^{m}$ at first order in $\th$. Using
$T_{\a\b}{}^c=-i(\C^c)_{\a\b}$, $T_{\a b}{}^c=0$ and a further choice
of gauge one finds

\beq
E_{\a}{}^m={i\over2}(\C^a)_{\a\b}e_a{}^m\th^{\n}\d_{\n}{}^{\b}+O(\th^2)\
. \la{4ealm} \eeq

Using the relation between the supervielbein and its inverse we can
then solve for $E_{\m}{}^a$ and then $E_a{}^m$ at first order in $\th$.
The connection is therefore given by

\beq \o_{a,bc}=\stackrel{o}{\o}_{a,bc} +Y_{a,bc} \la{4Y} \eeq

where $\stackrel{o}{\o}$ is the Levi-Civita connection and

\beq Y_{a,bc}= k_{a,bc} +k_{b,ac}-k_{c,ab} \la{4Y2} \eeq

with

\beq k_{a,bc}=-{i\over2}\left(\l_b{}^{\b'}(\C_a)_{\b'\c'}\l_c{}^{\c'}+
\psi_b{}^{\b}(\C_a)_{\b\c}\psi_c{}^{\c}\right) \ . \la{4k} \eeq

The terms with $\l$ vanish in the scalar equation of motion due to
symmetry and the spin-half equation of motion. We therefore find

\beq \h^{ab}\stackrel{o}{\nab}_a{\cE}_b{}^{\und}=
i\h^{ab}\left(\psi_a{}^{\a}(\C^{c'})_{\a\b'}\l_b{}^{\b'}
v_{c'}{}^{\und}-\psi_a{}^{\a}(\C_b)_{\a\b}\psi^{c\b}v_c{}^{\und}
+\ldots\right) \la{4kge3} \eeq

where the omitted terms are due to the target space connection and
four-form field strength. It is now straightforward to show that this
can be rewritten in the desired form, namely

\beq \h^{ab}\hat\nab_a{\cE}_b{}^{\und}=
-{1\over6}\e^{abc}{\cE}_c{}^{\unC}{\cE}_b{}^{\unB}{\cE}_a{}^{\unA}
G_{\unA\unB\unC}{}^{\und} \la{4kge4} \eeq

where the covariant derivative is the Levi-Civita derivative with an
additional term in the connection arising from the pull-back of the
target space connection, and where

\beq {\cE}_a{}^{\unA}=e_a{}^m\del_m z^{\unM} E_{\unM}{}^{\unA}|
\la{4emA} \eeq

The bilinear fermion terms in \eq{4kge4} arise from the dimension zero
component of $G_4$, namely, $G_{\ua\ub
\unc\und}=-i(\C_{\unc\und})_{\ua\ub}$.


\subsection{Kappa symmetry}


The basic embedding condition \eq{1ec} is intimately related to
$\k$-symmetry in the GS approach to branes
\cite{Sorokin:1989zi,Sorokin:1988nj}. Under an infinitesimal
worldvolume diffeomorphism one has

\be (\d z^{\unM})E_{\unM}{}^{\unA}=v^A E_{A}{}^{\unA} \label{3}\ , \ee

where $v^A$ is the worldvolume vector field generating the
diffeomorphism. For an odd diffeomorphism, with $v^a=0$, one finds,
using the embedding condition,

\be \d z^{\una}\equiv (\d z^{\unM})E_{\unM}{}^{\una}=0\ , \label{4} \ee

and

\be \d z^{\ua}\equiv (\d z^{\unM})E_{\unM}{}^{\ua}= v^{\a}
E_{\a}{}^{\ua}\ . \label{4a} \ee

This can be rewritten in the more familiar $\k$-symmetry  form

\be \d z^{\ua}= {1\over2}\k^{\ub}(1+\C)_{\ub}{}^{\ua}\ , \label{4b} \ee

where

\be \k^{\ua}=v^{\a} E_{\a}{}^{\ua}\ \label{4c} \ee

and where

\be P_{\ua}{}^{\ub}={1\over2}(1+\C)_{\ua}{}^{\ub} \label{4d} \ee

is the projection operator onto the odd tangent space of the
worldvolume from the odd tangent space of the target \footnote{$P$ and
$\C$ defined in the previous subsection are the $\th=0$ components of
corresponding objects in this subsection.}. It is given in terms of
$E_\a{}^{\ua}$ by

\be P_{\ua}{}^{\ub}=(E^{-1})_{\ua}{}^{\c} E_{\c}{}^{\ub}\ . \label{4e}
\ee

Thus we have

\bea
\d z^{\una}&=&0\nn\\
\d z^{\ua}&=&{1\over2}\k^{\ub}(1+\C)_{\ub}{}^{\ua} \ . \label{4f} \eea

Equations \eq{4f}, evaluated at $\th=0$, are the standard $\k$-symmetry
transformations of $z^{\unM}(x)$ in the GS formalism. The explicit form
of the operator $\C$, which must square to unity in order for $P$ to be
a projector, and the explicit relation between the parameters for
$\k$-symmetry and worldvolume supersymmetry depend on the choice of
basis for the odd tangent space on the worldvolume, but whichever basis
one chooses to work with, $\k$-symmetry has a precise definition in
terms of worldvolume supergeometry. For our choice of basis it is easy
to see, using Appendix A, that

\bea \C_{\ua}{}^{\ub}&=&-{1\over6}\e^{abc}(\C_{abc})_{\ua}{}^{\ub}\nn
\w2
&\rightarrow&\d_\a{}^\b\left(\barr{cc} \d_i{}^j & 0\\
0 & -\d_{i'}{}^{j'}\earr\right) \eea

so that

\be
P_{\ua}{}^{\ub}\rightarrow   \d_\a{}^\b\left(\barr{cc} \d_i{}^j & 0\\
0 & 0\earr\right) \ee

as required.


\subsection{The action formula}


In this subsection we briefly review the action formula for arbitrary
$p$-branes given in \cite{Howe:1998ts}. We shall then apply it to the
M2-brane. For any brane we assume the existence of a closed Wess-Zumino
form $W_{p+2}$ defined on the worldvolume $M$; it is given  by

\be W_{p+2}=d Z_{p+1} \la{4Z} \ee

where $Z$ is a potential $(p+1)$-form. Since $W$ it is a $p+2$-form on
a manifold which has even (i.e. bosonic) dimension $p+1$ it follows
that it is exact. This is so because the de Rham cohomology of a
supermanifold coincides with the de Rham cohomology of its body.
Therefore we can always write it in the form

\be W_{p+2}= d K_{p+1} \label{7} \ee

for some {\it globally} defined $(p+1)$-form $K$ on $M$. Furthermore,
since none of the target space fields or the worldvolume fields has
negative dimension, at least at lowest order in derivatives, it follows
that the only non-vanishing component of $K$ is the purely bosonic one.
In components this means

\be K_{\a A_1 \cdots A_p}=0\ . \label{7.1} \ee

We now define the Green-Schwarz Lagrangian form $L_{p+1}$ to be

\be L_{p+1}= K_{p+1}-Z_{p+1} \ . \label{8} \ee

Under a worldvolume diffeomorphism generated by the vector field $v$
one has

\be \d L_{p+1}={\cal L}_v L_{p+1} =d i_v L_{p+1} + i_v d L_{p+1}
\label{9} \ee

Since, by construction, $L_{p+1}$ is closed, the variation \eq{9}
reduces to

\be \d L_{p+1}=d i_v L_{p+1}\ . \la{9a} \ee

Therefore the action integral

\be S=\int_{M_0}\, L^0_{p+1}\ , \la{9b} \ee

where $M_0$ is the body of $M$ and where

\be L^0_{p+1}=dx^{m_{p+1}}\wedge dx^{m_p} \wedge \ldots dx^{m_1}
L_{m_1\ldots m_{p+1}}\vert\ , \la{9c} \ee

is invariant under $\k$-symmetry transformations and diffeomorphisms of
$M_0$, since these transformations are identified with the leading
components of the superdiffeomorphisms of $M$. The use of closed
superspace $d$-forms to construct superinvariants in $d$ spacetime
dimensions has been noted elsewhere in the literature
\cite{D'Auria:1982pm,Gates:1997ag}.

The Wess-Zumino form for the M2-brane is the pull-back of the
supergravity four-form $G_4$. On the worldvolume of the brane there
should therefore be a three-form $K_3$ such that

\be W_4=f^* G_4= d K_3\ . \la{9o} \ee

In index notation this reads

\be 4\nab_{[A}K_{BCD]}+ 6T_{[AB}{}^E K_{|E|CD]}= (f^*G)_{ABCD}\ .
\la{9p} \ee

Since there are no fields of negative dimension on the worldvolume
(given the standard embedding condition), the only non-vanishing
component of $K$ has purely even indices. By directly evaluating the
dimension zero component of the \eq{9o} one finds that it is satisfied
for

\be K_{abc}=\e_{abc}\ . \la{9q} \ee

It is then easy to verify that the entirety of \eq{9o} is satisfied
\cite{Howe:1998ts}. In a coordinate basis one has

\be K_{mnp}|=\e_{mnp} \sqrt{-\det g}\ , \la{9s} \ee

where $g$ is the GS metric. The GS Lagrangian is therefore recovered
from the general formulae \eq{8} and \eq{9b}; it is \cite{Howe:1998ts}

\be {\cL}=\sqrt{-\det g} -{1\over 6}\e^{mnp}\del_p z^{\unP}\ \del_n
z^{\unN}\ \del_m z^{\unM}\ C_{\unM\unN\unP}\ , \la{9t} \ee

where $G_4=dC_3$ on $\unM$, and where

\be L^0=dx^m\wedge dx^n\wedge dx^p \e_{mnp}\ {\cL}\ . \la{9u} \ee


\section{The induced supergeometry}


In this subsection it will be shown that the  geometry induced on the
worldvolume, discussed briefly in \cite{Bandos:1995zw}, is that of
$d=3,\ N=8$ conformal supergravity. The induced multiplet is composite.
This will be done by comparing the worldvolume geometry to the known
superspace geometry of $d=3,\ N=8$ conformal supergravity
\cite{Howe:1995zm}.  In order to make the comparison it is convenient
to redefine the worldvolume geometry such that the torsion constraints
are in standard form. To simplify matters we shall take the target
space to be flat $(11|32)$-dimensional superspace as the presence of
non-trivial target space terms does not affect the structure of the
worldvolume supergravity multiplet.


\subsection{The $d=3,N=8$ conformal supergravity multiplet}


We shall use tildes to distinguish worldvolume geometrical objects in
standard form from the induced form partially listed in the preceding
subsection. Since the dimension zero and one-half components of the
induced torsion are already in standard form, the only amendments that
need to be made are at dimension one. If we set

\beq \tO_{a,b}{}^{c}=\O_{a,b}{}^{c} + M_{a,b}{}^c \la{5conn} \eeq

where

\beq M_{a,bc}=-{i\over2}(\bar\L_a\C_b\L_c-\bar\L_a\C_c\L_b
+\bar\L_b\C_a\L_c) \la{5M} \eeq

we find that

\beq \tT_{ab}{}^c=0 \la{5d1t} \eeq

and

\beq \tT_{a\b}{}^{\c}\rightarrow\tT_{a\b j}{}^{\c
k}=\d_j{}^k(\c_a)_{\b}{}^{\c}U + (\c^b)_{\b}{}^{\c} V_{abj}{}^{k}
\la{5d1t2} \eeq

where

\beqa U&=& {i\over8}\bar\L^{a i'}\L_{a i'} \nonumber\w2 V_{ab
ij}&=&-{i\over8}(\s_{a'b'})_{ij}(\s^{a'b'})_{i'j'}\bar\L_a^{i'}\L_b^{j'}\
. \la{5U} \eeqa

The dimension one torsion is now in standard form (for a flat target
space) so that the redefinition \eq{5conn} is the only one that needs
to be made in this case. The non-zero dimension one torsion for $d=3,\
N=8$ conformal supergravity has the same form as \eq{5d1t2}, except
that $\d_{jk}U$ is replace by $U_{jk}$ where the latter is a symmetric
matrix of scalar fields. As we shall see in the next section, the
particular form of $U$ that occurs here is a consequence of the choice
of conformal gauge that we have made for the superembedding. However,
neither of the leading components of $U$ or $V$ are fields in the
conformal supergravity multiplet. To exhibit such a field it is
necessary to compute the dimension one curvature $R_{\a i\b j,kl}$. The
field redefinition \eq{5conn} does not change this component of the
curvature so it can be calculated from the Gauss-Codazzi equations
using our previous results. We find

\beqa \tR_{\a i\b j,kl}& =&
i\e_{\a\b}\left(W_{ijkl}+2(\d_{ik}\d_{jl}-\d_{jk}\d_{il})U\right)
\nonumber\w2 &&
+i(\c_a)_{\a\b}\left(\d_{ij}V^a_{kl}-4\d_{(i[k}V^a_{j)l]}\right)
\la{5R} \eeqa

where

\beq W_{ijkl}={i\over64}(\s^{a'b'})_{ij}(\s^{c'd'})_{kl}
(\s_{a'b'c'd'})_{i'j'}\bar\L^{a i'}\L^{j'}_a \ . \la{5W} \eeq

This field is totally antisymmetric and anti-self-dual on the $SO(8)$
indices and is part of the conformal supergravity multiplet. One might
consider it to be the super-Weyl tensor for $d=3,\ N=8$ supergeometry.

There are additional, higher-dimensional fields in the supergravity
multiplet but we shall not go into further details here. However, the
calculations presented above show that the induced multiplet is indeed
a conformal supergravity multiplet as claimed. The components of this
multiplet are (in order of increasing dimension): $(e_m{}^a;\
\psi_m^i;\ A_{mij},\,B_{ijkl};\ \r_{ijk};\ C_{ijkl})$, where each field
is antisymmetric on its internal indices.  As usual, $e$ is the
graviton and $\psi$ represents the gravitini, while $A$ is the $SO(8)$
gauge field, $B$ and $C$ are scalar fields (anti-self-dual) and $\r$ a
set of spinor fields. The field $B$ is the leading component of the
superfield $W$ so that it is the lowest-dimensional non-gauge field in
the multiplet. Not counting gauge degrees of freedom one sees that this
multiplet has 128 + 128 degrees of freedom.


\subsection{Super-Weyl transformations}


We shall briefly discuss the effect of super-Weyl transformations on
the worldvolume geometry since this establishes the fact that $U$ and
$V$ do not belong to the conformal supergravity multiplet. We shall
work with the standard geometry (marked by a tilde) and mark the
transformed quantities by hats. The computations are most easily
carried out using differential forms. The super-Weyl transformations
are, for the basis forms,

\beqa \tE^a\rightarrow\hat E^a &=& e^{-2S}\tE^a \nonumber\w2
\tE^{\a}\rightarrow\hE^{\a}&=&e^{-S}\big(\tE^{\a} +\tE^a
\S_a{}^{\a}\big) \la{5sw} \eeqa

where

\beq \S_a{}^{\a}=-2i(\C^a)^{\a\b}\tnab_{\b}S\ . \la{5sig} \eeq

Dually, for the basis vectors, we have

\beqa \hE_{\a}&=&e^S\tE_{\a}\nonumber\w2 \hE_a&=& e^{2S}\big(\tE_a
-\S_a{}^{\a}\tE_{\a} \big) \la{5sw2} \eeqa

For the connection one has

\beq \hO_{A}{}^{B}=\tO_{A}{}^{B} +Y_A{}^B \la{5sw3} \eeq

where, at dimension one-half,

\beqa Y_{\a i,bc}&=& 2(\c_{bc})_{\a}{}^{\b}\S_{\b i}\nonumber\w2 Y_{\a
i,jk}&=&2(\d_{ik}\S_{\a j}-\d_{ij}\S_{\a k}) \la{5Y} \eeqa

with $\S_{\a i}:=\tnab_{\a i}S$. It follows that

\beq Y_{\a i,\b j\c k}=\d_{jk}(\e_{\b\a}\S_{\c i}+\e_{\c\a}\S_{\b
i})+2\e_{\b\c}(\d_{ik}\S_{\a j}-\d_{ij}\S_{\a k})\ . \la{5Y2} \eeq

At dimension one,

\beqa Y_{a,bc}&=& 2(\h_{ab}\tnab_c S-\h_{ac}\tnab_b S)
+2\bar\S^i\S_i\nonumber\w2 Y_{a,ij}&=& L_{a ij} + i\bar\S_i\c_a\S_j
\la{5Y3} \eeqa

where

\beq \tnab_{\a i}\tnab_{\b j}S==i\big(\e_{\a\b}K_{ij}
+(\c^a)_{\a\b}L_{a ij}+{i\over2}\d_{ij}(\c^a)_{\a\b}\tnab_a S\big) \ .
\la{5DDS} \eeq

This equation should be read as defining the fields $K_{ij}$ and
$L_{aij}$ as components of the scalar superfield $S$. The former is
symmetric while the latter is antisymmetric as $SO(8)$ tensors.

On substituting the above transformations into the expressions for the
torsion and curvature with the standard constraints one finds that all
the components of the torsion of dimension less or equal to one are
unaffected except for the component $T_{a\b}{}^{\c}$ for which one
finds

\beq \hT_{a\b j}{}^{\c k}=e^{2S}\big(\tT_{a\b j}{}^{\c k} +
2(\c_a)_{\b}{}^{\c}K'_j{}^k +
2\e_a{}^{bc}(\c_b)_{\b}{}^{\c}L'_{cj}{}^k\big) \la{5d1t3} \eeq

where

\beqa K'_{ij}&=&K_{ij}-i\bar\S_i\S_j +
{i\over2}\d_{ij}\bar\S^k\S_k\nonumber\w2
L'_{aij}&=&L_{aij}-i\bar\S_i\c_a\S_j \ . \la{5defs} \eeqa

Thus, as expected, in a general conformal gauge, the component of the
torsion tensor given in \eq{5d1t3} is made up from a symmetric scalar
and an antisymmetric vector. The result also shows that the leading
components of $U$ and $V$ in \eq{5d1t2} can be gauged to zero by an
appropriate choice of the $\th^2$ parameter in $S$. For the dimension
one curvature we find

\beq \hR_{\a i\b j,kl}=e^{2S}\big(\tR_{\a i\b j,kl}
+8i\e_{\a\b}\d_{[i[k}K'_{j]l]}
+2i(\c_a)_{\a\b}(\d_{ij}L'^a_{kl}-4\d_{(i[k}L'^a_{j)l]})\big)\ .
\la{5R2} \eeq

This equation confirms that $W_{ijkl}$ can indeed be viewed as the
super-Weyl tensor as it simply scales under super-Weyl transformations.


\section{The gauged M2-brane}


In this section we shall discuss gauged superembeddings for target
superspaces which admit isometries. We shall focus on the case when
there is just one $U(1)$ isometry, but the formalism can be
straightforwardly generalised to the non-abelian isometries of group
manifolds.\footnote{Non-abelian gauging in the Green-Schwarz model has been
considered in \cite{Nishino:2003yn}.} In effect, this is a gauged sigma model in
a superspace setting, but without a kinetic term for the gauge field associated
with the isometry. As we shall see, this gauge field turns out to be the
pull-back of the type IIA RR 1-form, but the worldvolume Born-Infeld gauge field
appears in a natural manner in the action formula. While the $U(1)$ gauging of
the supermembrane was discussed in \cite{Nishino:2003yn} the role played by the
$U(1)$ gauge field was apparently not fully appreciated there.


\subsection{Gauged M2-brane superembedding}


We consider the superembedding $f:M\rightarrow \huM$, from the
worldvolume $M$ into the $D=11$ target superspace. We denote this by a
hat as we shall use $\unM$ for the $D=10$ IIA superspace, the two being
related by the projection $\unp:\huM\rightarrow \unM$. The $U(1)$
isometry is generated by a vector field $\huK$ so that geometrical
quantities will be annihilated by the Lie derivative along $\huK$. The
gauge superembedding matrix is

\be D f:=d f + C_1 \otimes \huK \la{6df} \ee

where $C_1$ is the $U(1)$ gauge field on $M$. We denote the components
of $D f$ with respect to standard bases on both superspaces by
$\tE_A{}^{\huA}$. We have

\be \tE_A{}^{\huA}=E_A{}^M(\del_M z^{\huM}+C_M
K^{\huM})E_{\huM}{}^{\huA}\ . \la{6em} \ee

It is convenient to follow the standard practice  and work with the
gauge-covariant pull-back defined by using $D f$ in place of $d f$. In
other words, in a coordinate basis, $\del_M z^{\huM}$ is replaced by
$\del_M z^{\huM}+C_M K^{\huM}$. Given a target-space form $\huo$ the
gauge-covariant pull-back $\tf \huo$ can be written

\be \tf \huo=f^*\huo +f^*(i_{\huK}\huo)\wedge C_1\ . \la{6gc} \ee

A useful identity is

\be d\tf\huo=\tf d\huo+G_2\wedge\tf i_{\huK}\huo
+f^*\cL_{\huK}\huo\wedge C_1 \la{6ui} \ee

where $G_2=d C_1$. The last term vanishes on invariant forms and so
plays no r\^ole in the following considerations.

The torsion equation is modified slightly:

\be 2\hat{\nab}_{[A} \tE_{B]}{}^{\huC} +
T_{AB}{}^{C}\tE_C{}^{\huC}+G_{AB}K^{\huC}=
\tE_{B}{}^{\huB}\tE_{A}{}^{\huA}\hT_{\huA\huB}{}^{\huC}, \la{6tor} \ee

where $G_2=dC_1$. The covariant derivative $\hat{\nab}$ in this formula
involves a worldvolume connection acting on the worldvolume index of
$\tE_A{}^{\huA}$ and the covariant pull-back of the target space
connection acting on the target space index. It does not involve a
separate term  with the $U(1)$ gauge field because the embedding matrix
is $U(1)$-invariant.

We choose the embedding condition to be the gauged version of the usual
one, namely

\be \tE_{\a}{}^{\hat{\una}}=0 \la{6ec} \ee

We can also parametrise the non-zero components of $\tE_A{}^{\huA}$ in
exactly the same way as the usual membrane with the aid of an element
$\hu$ of the eleven-dimensional Lorentz group. However, we are not
allowed to set the field $h_{\a}{}^{\b'}$ (introduced in \eq{1h}) to
zero as the torsion equation has been modified. To make further
progress we make a $10+1$ split of the even target space quantities. We
put $z^{\huM}=(z^{\unM},y)$ and
$E^{\huA}=(\hat{E}^{\unA},\hat{E}^{11})$\footnote{The KK reduction of
the supervielbein is described in Appendix C.}. The $A=\a,B=\b,\huC=11$
component of the torsion equation reads

\be T_{\a\b}{}^{c} \tE_{c}{}^{11}+G_{\a\b} \huK^{11}=
\tE_{\a}{}^{\ua}\tE_{\b}{}^{\ub}\hT_{\ua\ub}{}^{11} \la{6t11} \ee

At this stage we note that we are free to impose a conventional
constraint on $G_{\a\b}$ by shifting the even part of $C$. The natural
choice is to do this in such a way that

\be T_{\a\b}{}^{c} \tE_{c}{}^{11}=0 \la{6e11} \ee

and this in turn implies that

\be \tE_A{}^{11}=0\ . \la{611} \ee

This implies that, up to a gauge transformation, the worldvolume gauge
field $C_1$ is the pull-back of the IIA target space RR 1-form
$\unC_1$,

\be C_1=f_o^* \unC_1 \la{6c1} \ee

where $f_o:=\unp\circ f$ is the IIA superembedding. In components it is
given by $z^M \rightarrow z^{\unM}(z)$. With the aid of \eq{611} it is
simple to show that

 \be
\tE_A{}^{\unA}=E_A{}^{\unA}\ , \la{6ta}
 \ee

where $E_A{}^{\unA}$ is the IIA embedding matrix. The  $\huC=11$
component of the torsion equation now reads

 \be
 G_{AB} \huK^{11}=E_B{}^{\unB}E_A{}^{\unA}
 \left(\hT_{\unA\unB}{}^{11} -2\hat{\O}_{[\unA\unB]}{}^{11}\right)
\la{6gab1}
 \ee

The connection component that appears in this equation comes from the
$\hat{\nab}$ covariant derivative in the torsion equation \eq{6tor}. As
can be seen from \eq{KK3.1} the expression in brackets on the
right-hand-side of \eq{6gab1} is the field strength tensor of the RR
1-form $\unC$ multiplied by the scalar field $\F$. Since $\huK^{11}=\F$
we see that the field strength of the $U(1)$ gauge field is identified
with the pull-back of the RR 2-form field strength as expected. The
$\unC$ component of the torsion equation now becomes the usual torsion
equation of the D2-brane,

 \be
2{\nab}_{[A} E_{B]}{}^{\unC} + T_{AB}{}^{C}E_C{}^{\unC}=E_{B}{}^{\unB}
E_{A}{}^{\unA}T_{\unA\unB}{}^{\unC}\ . \la{6tor2}
 \ee

In deriving this equation one has to use the fact that additional terms
present in \eq{6tor} cancel on using the Kaluza-Klein reduction of the
$D=11$ superspace geometry given in appendix C.


\subsection{Geometrical structure, Wess-Zumino form and action}


The geometry of the preceding subsection can be clarified by the
introduction of  a $U(1)$ bundle $\hat M$ over $M$. We can define a
gauged superembedding to be an equivariant map $\hat f:\hat
M\rightarrow \hat {\unM}$. Thus we have the following picture:

\begin{center}
\unitlength=1mm
\begin{picture}(100,50)
\put(15,35){$\hat M$} \put(75,35){$\hat{\unM}$} \put(15,5){$M$}
\put(75,5){$\unM$} \put(25,6){\vector(1,0){45}}
\put(25,36){\vector(1,0){45}} \put(15,30){\vector(0,-1){20}}
\put(20,10){\vector(0,1){20}} \put(77,30){\vector(0,-1){20}}
\put(25,10){\vector(2,1){45}} \put(45,40){$\hat f$} \put(45,10){$f_o$}
\put(10,20){$p$} \put(22,20){$s$} \put(80,20){$\unp$} \put(45,27){$f$}
\end{picture}
\end{center}

Here $p:\hat M\rightarrow M$ and $\unp:\hat{\unM}\rightarrow \unM$ are
the projections, $s:M\rightarrow\hat M$ is a section of $\hat M$,
$f_o:M\rightarrow\unM$ is the IIA superembedding, $\hat f:\hat
M\rightarrow\hat{\unM}$ is the lifted map and
$f:M\rightarrow\hat{\unM}$ is a gauged superembedding representative.
The diagram is commutative so that the maps are related to each other
in ways that can be read off straightforwardly. The equivariance of
$\hat f$ means that

\be \hat f \circ g= g\circ \hat f \la{6circ} \ee

where $g$ denotes an element of $U(1)$ group acting on $\hat M$ on the
left-hand side and $\hat{\unM}$ on the right-hand side. From the
diagram we have:

\bea
f&=& \hat f\circ s\nn\\
f_o&=&\unp\circ f=\unp\circ\hat f\circ s \ . \la{6circ2} \eea

It should be emphasised that the above diagram and discussion is local.
In general, we might have non-trivial bundles in which case neither $s$
nor $f$ is globally defined. In this case the equivariance of the map
$\hat f$ still allows us to define $f_o$ uniquely from

 \be
 f_o\circ p= \unp\circ \hat f\ .
 \la{6circ3}
 \ee

Another way to think about the problem is that, given an embedding
$f_o:M\rightarrow \unM$ in IIA such that there is a $U(1)$ bundle
$\hat{\unM}\rightarrow\unM$, then we can define $\hat M\rightarrow M$
as the pull-back of $\hat{\unM}$. The maps $\hat f$, $s$ and $f$ can
then be defined such that we obtain the above diagram. It is then
natural to identify the connection form $\ho$ of $\hat M$ with the
pull-back under $\hat f$ of the connection form $\huo$ of $\hat{\unM}$.

The above discussion makes it clear that the gauged supermembrane in
$D=11$ is simply a lifted version of the D2-brane in IIA. However, the
formalism gives us a neat way to derive the D2 brane action directly
starting from the generalised Wess-Zumino term without the need to
dualise the additional transverse coordinate $y$ to a worldvolume
Born-Infeld field.

As we shall see below, this construction requires the introduction of a
field-strength 2-form $\cF_2$ on the worldvolume which obeys a modified
Bianchi identity. This is the field strength of the gauge field on the
brane. As the Wess-Zumino 4-form which we derive this way is exactly
the same as the one obtained from studying the D2-brane directly in ten
dimensions, it follows, provided we impose the usual embedding
condition on $f_o$, that the brane action principle will lead us to the
usual Green-Schwarz action complete with the Dirac-Born-Infeld kinetic
term and the D-brane Wess-Zumino term.

Let $\huG_4$ denote the closed 4-form of $D=11$ supergeometry on
$\huM$, and let $\hG_4$ denote its pull-back to $\hM$. We shall assume
that $\huG_4$ is $U(1)$ invariant so that

\be d\huG_4=\cL_{\huK}\huG_4=0 \ . \la{6hug} \ee

On the other hand $i_{\huK}\huG_4 =\unH_3$, the 3-form of type IIA. On
$\hM$, the pull-back $\hG_4$ is again invariant and closed, and $i_K
\hG_4$ is closed, invariant and basic, that is, it satisfies
(trivially) $i_K(i_K \hG_4))=0$. It can therefore be expressed in terms
of a 2-form $\cF_2$ on $\hM$ which we can also assume to be invariant
and basic. So we have

\be i_K\hG_4=-d\cF_2 \la{6f} \ee

where

\be \cL_K\cF_2=i_K\cF_2=0 \ . \la{6f2} \ee

We would like to construct a Wess-Zumino 4-form $W_4$ on $M$ in order
to apply the brane action principle. The construction is essentially
the same as that of  \cite{Hull:1990ms} (see \cite{Percacci:1998ag} for
a review) but differs slightly in that we shall work on $\hM$ rather
than on $M$. In our approach a gauge-invariant Wess-Zumino form is
equivalent to a closed, invariant, basic form $\hW_4$ on $\hM$ with
$\hW_4=p^*W_4$. Now $\hG_4$ is not basic, but we can correct this by
adding to it a term of the form $\o\wedge i_K G_4$ where $\o$ is the
connection 1-form on $\hM$. This works because the connection form
satisfies $i_K\o=1$. The modified 4-form is no longer closed, but this
in turn can be corrected by adding an extra term of the form
$F_{\o}\wedge\cF_2$, where $F_{\o}=d\o=p^*G_2$ is the curvature 2-form.
We thus arrive at a lifted Wess-Zumino 4-form

\be \hW_4=\hG_4+ \o\wedge i_K G_4 + F_{\o}\wedge\cF_2 \la{6wz} \ee

On $\huM$ we have

\be \huG_4=\unG_4 + \huo\wedge \unH_3 \la{6hug2} \ee

where $\huo$ is the connection 1-form of $\huM$. In standard
coordinates this can be written $\huo=d\uny+ \unC_1$ where $\unC_1$ is
the RR 1-form. Using this we find that $d\cF_2=-H_3:=-f_o^*\unH_3$ and
so

\be W_4=G_4 + G_2\cF_2 \la{6wz2} \ee

where $G_2=f_o^*\unG_2=f_o^*(d\unC_1)$. This is the standard
Wess-Zumino form for the D2-brane. The corresponding potential is

\be Z_3=f_o^*\unC_3 + f_o^*\unC_1\wedge\cF_2 \la{6z} \ee

The usual Green-Schwarz action for the D2-brane can easily be
reconstructed starting from $W_4$ with the aid of the brane action
formula described in $4.4$. This was done in detail in
\cite{Howe:1998ts}; for the reader's convenience we outline the main
steps.

We parametrise the dimension zero components of the embedding matrix in
the form \cite{Howe:1996yn}

\bea
E_{\a}{}^{\ua} &=& u_{\a}{}^{\ua} + h_{\a}{}^{\b'} u_{\b'}{}^{\ua} \nn\\
E_{a}{}^{\una}&=& u_a{}^{\una}\ . \label{10} \eea

Here $u$ denotes part of a matrix of the group $Spin(1,9)$, in the
spinor or the vector representations according to the indices. An
analysis of the dimension zero component of the torsion equation
yields, on using the usual embedding condition $E_{\a}{}^{\una}=0$,

\bea h_{\a}{}^{\b'}\rightarrow h_{\a i}{}^{\b' j} &=&
\d_i{}^j(\c^{ab})_{\a}{}^{\b'} h_{ab}\w2 T_{\a\b}{}^c&=&-i(\C^d)_{\a\b}
m_d{}^c\ , \label{15} \eea

where

\be m_a{}^b=\d_a{}^b(1-4y) + 8(h^2)_a{}^b ,\qquad y:={1\over 2} {\rm
tr}\, h^2 \label{17} \ee

and where $h^2$ denotes matrix multiplication. Using these results, one
can show that the $\cF$ identity, $d\cF=-f_o^*H_3$, is satisfied
provided that we choose all the components of $\cF$ to vanish except
for $\cF_{ab}$ which is related to $h$ by

\be m_a{}^c \cF_{cb}= 4h_{ab}\ . \label{19} \ee

This can be rearranged to give

\be \cF_{ab}={4 h_{ab}\over 1+4y}\ . \la{fh} \ee

The Wess-Zumino four-form, $W_4=d(f^*C_3 + f^*C_1 \cF)$, can be
rewritten as $dK_3$. From the above constraints we find that all of the
components of $K_3$ vanish except for $K_{abc}$ which is given by

\be K_{abc}=\e_{abc}\ K\ , \ee

with

\be K={1-4y\over 1+4y}\ . \label{25} \ee

$K_3$ is the same as for the M2-brane case except for the factor $K$ in
$K_{abc}$. As shown in \cite{Howe:1998ts} the above expression \eq{25}
for $K$ can be rewritten as

\be K=\sqrt{{\rm det}(\d_m{}^n + \cF_m{}^n)}\ . \label{30} \ee

The GS action is obtained by integrating the bosonic part of the closed
3-form $L_3=K_3-Z_3$ denoted by $L^0$. The Lagrangian is

\be L^0=dx^m\wedge dx^n\wedge dx^p \e_{mnp}\ {\cL}\ . \la{9uu} \ee

with

\be {\cL}= \sqrt{-{\rm det}(g_{mn} + \cF_{mn})} -f^*C_3 - f^*C_1 \cF\ .
\ee

This form of the Lagrangian is given in what we have called the KK
frame in Appendix C. In the string frame the kinetic term in  the
action acquires a factor of $e^{-\f}$.

\section{Massive branes}


Massive D-branes are D-branes whose worldvolume theories include a
massive Chern-Simons term for the worldvolume gauge field. They were
discussed in \cite{Bergshoeff:1997cf} where it was shown that
kappa-symmetry requires the target space to be that of the Romans type
IIA supergravity. Subsequently, it was shown how the massive D2-brane
could be lifted to eleven dimensions, at least in components. In this
section we shall discuss this lifting in the superembedding framework.
In the next subsection we omit underlines and hats from
eleven-dimensional quantities as the discussion does not involve any
branes.


\subsection{Massive $D=11$ supergeometry}

In \cite{Bergshoeff:1997ak} it was shown how one rewrite massive IIA
supergravity in eleven dimensions, although the resulting theory is not
conventional $D=11$ supergravity since one has to introduce a Killing
vector which cannot be eliminated. Here we apply a similar idea in the
superspace context. Some partial results were obtained on this in
\cite{Nishino:2003yn}.

The data in eleven dimensions is a vector field $K$ which generates the
action of a circle group, a 3-form gauge field $C_3$ and a set of
1-forms $E^A$. All the fields we need to consider are invariant forms
$\o$, so that $\cL_K \o=0$. We can define a transformation on such
forms in the following way:

\be \d \o=m i_K\o\wedge i_K \L_2 \la{7do} \ee

where the 2-form parameter $\L_2$ is itself invariant. A covariant
exterior derivative can be defined with the aid of the  3-form gauge
field $C_3$. If

\be \d C_3=d\L_2 + m i_K C_3 \wedge i_K \L_2\ , \la{7dc3} \ee

then it is easy to check that

\be D\o:=d\o + {m} i_K \o\wedge i_K C_3 \la{7cov} \ee

transforms in the same way as $\o$. The curvature 4-form $G_4$ is
defined by

\be G_4=d C_3 + {m\over2} i_K C_3 i_K C_3 \la{7curv} \ee

and is easily seen to transform covariantly.

The basis 1-forms $E^A$ transform in the usual way under local Lorentz
transformations. Since $E^A$ is invariant under $K$, the Lorentz
transformation parameter must be as well. $E^A$ is taken to transform
under $\L_2$ transformations as a 1-form so that the definition of the
torsion needs to be modified as follows:

\be T^A= D E^A= dE^A +  E^B\O_B{}^A + m i_K E^A i_K C_3 \la{7tor} \ee

This definition is similar to the one given in \cite{Bergshoeff:1997ak}
for the bosonic part of the geometry, but with the difference that the
extra term cannot be absorbed into the Lorentz connection as it does
not have the right symmetry properties on the whole of superspace. The
Bianchi identities which follow from this definition are

\be D T^A =E^B R_B{}^A -m i_K E^A i_K G_4 \la{7tb} \ee

where the curvature is given by

\be R_A{}^B=d\O_A{}^B + \O_A{}^C \O_C{}^B +mi_K\O_A{}^B i_K C_3\ .
\la{7R} \ee

Note that this curvature is Lorentz-valued and is covariant under both
Lorentz and $\L_2$ transformations. The Bianchi identity for the 4-form
field strength is simply

\be DG_4=0\ . \la{7gb} \ee

The constraints at dimension less than one on the curvatures for this
geometry are the same as in the massless case, i.e. the only
non-vanishing components of $T$ and $G$ are

\bea
T_{\a\b}{}^c&=&-i(\C^c)_{\a\b} \\
G_{\a\b cd}&=&-i(\C_{cd})_{\a\b} \eea

At dimension one we have the field-strength $G_{abcd}$ which appears in
the dimension one torsion $T_{a\b}{}^{\c}$ as before, but there are
also mass-dependent corrections. The result is

\be T_{a\b}{}^{\c}=\bar T_{a\b}{}^{\c} + {m\over12}\left( K^2
(\C^c)_{\a\b} - 6m K_a K^b (\C_b)_{\a\b}\right) \ee

where $K^2=K^a K_a$ and $\bar T_{a\b}{}^{\c}$, given in \eq{3td1},
corresponds to the massless case. This can be obtained by examining the
dimension one Bianchi identities \eq{7tb}. The mass-dependent terms
originate from the second term on the right-hand side of \eq{7tb}.


\subsection{The massive gauged superembedding}


We shall now consider a gauged superembedding $f:M \rightarrow \huM$
where $\huM$ is the massive $D=11$ superspace described above. We
define the gauged superembedding matrix as in the massless case.
However, the presence of the mass term necessitates a further
modification to the torsion identity which becomes

\be 2\hat{\nab}_{[A} \tE_{B]}{}^{\huC} +
T_{AB}{}^{C}\tE_C{}^{\huC}+\cG_{AB}K^{\huC}=
\tE_{B}{}^{\huB}\tE_{A}{}^{\huA}\hT_{\huA\huB}{}^{\huC} \la{7tor2} \ee

where

\be \cG_2=G_2 + m\tf i_{\huK} \huC_3\ . \la{7cg} \ee

Note that this implies that

\be d\cG_2=m \tf i_{\huK} \huG_4\ . \la{7dcg} \ee

The only modification is in the change $G_2\rightarrow\cG_2$ which
comes from the covariant pull-back of the mass-dependent term in the
torsion. Note that the covariant derivative in \eq{7tor2} contains only
the induced Lorentz connection and the worldvolume connection. As usual
we impose the embedding condition

\be \tE_{\a}{}^{\hat{\una}}=0\ . \ee

Following a similar line of argument to the massless case, we make a
10+1 split of the even coordinates and choose the bosonic part of the
worldvolume gauge field $C_1$ such that the embedding trivialises in
the eleventh direction, i.e. $\tE_A{}^{11}=0$. This again implies that
$C_1=f_o^* \unC_1$ and that $\tE_A{}^{\unA}= E_A{}^{\unA}$. We now have
\eq{6gab1} but with $G_{AB}$ replaced by $\cG_{AB}$,

 \be
 \cG_{AB} K^{11}=E_B{}^{\unB}E_A{}^{\unA}
 \left(\hT_{\unA\unB}{}^{11} -2\hat{\O}_{[\unA\unB]}{}^{11}\right)\ .
\la{7gab}
 \ee

In this case it is the modified field strength of the $U(1)$ gauge
field which is identified with the pull-back of the RR 2-form field
strength. In other words

\be \cG_2=f_o^* \unG_2 \la{7gg} \ee

Finally the $\unC$ component of the torsion equation becomes the
torsion equation of the massive D2-brane,

\be 2{\nab}_{[A} E_{B]}{}^{\unC} +
T_{AB}{}^{C}E_C{}^{\unC}=E_{B}{}^{\unB}
E_{A}{}^{\unA}T_{\unA\unB}{}^{\unC}\ . \la{7tor3}
 \ee

This is formally identical to the massless case, but the target
superspace geometry is now that of massive IIA supergravity.


\subsection{The Wess-Zumino form }


We shall derive the massive D2-brane action starting from the covariant
pull-back of $\huG_4$. The idea is to add terms to $\tf\huG_4$ in order
to obtain a closed Wess-Zumino 4-form. Using \eq{6ui} we find

\be d\tf \huG_4= \cG_2\wedge \tf(i_{\huK} \huG_4) \la{7dfg} \ee

Using \eq{6ui} again we can easily show that $\tf(i_{\huK} \huG_4)$ is
closed, so that we can write

\be \tf(i_{\huK} \huG_4)=-d\cF_2 \la{7F2} \ee

where $\cF_2$ is a worldvolume 2-form which we identify with the field
strength of the D2-brane gauge field. We then have

\be d\tf \huG_4= -\cG_2\wedge d\cF_2 =-d(\cG_2\wedge\cF_2)
+d\cG_2\wedge\cF_2 \la{7id} \ee

Equation \eq{7dcg} enables us to replace $d\cG_2$ by $m\tf(i_{\huK}
\huG_4)$ and this in turn is equal to $-md\cF_2$. The second term on
the right of \eq{7id} is thus exact and we can therefore define the
Wess-Zumino 4-form $W_4$ to be

\be W_4=\tf \huG_4 +\cG_2\wedge\cF_2 + {m\over2} \cF\wedge\cF \la{7w4}
\ee

Substituting the 10+1 split into $W_4$, we see that the first term
reduces to $f_o^*\unG_4$,

\be \tf \huG_4=\tf\left(\unG_4 + E^{11}\wedge
(i_{\huK}\huG_4)\right)=f_o^*\unG_4 \la{7w41} \ee

as $\tf E^{11}=0$. Using the fact that $\cG_2=f_o^* \unG_2$ we get

\be W_4=f_o^* \unG_4 + \unG_2 \wedge\cF_2 +{m\over2} \cF_2\wedge\cF_2
\la{7w42} \ee

This is the standard massive D2-brane Wess-Zumino form which can be
written as $dZ_3$ where

\be Z_3=f_o^*\unC_3 +f_o^*\unC_1\wedge \cF_2 + {m\over2} AdA \ee

The GS Lagrangian can be obtained by following the same steps as in the
massless case. The only formal difference is the occurrence of the
Chern-Simons term for the D-brane gauge field. Thus we find

\be {\cL}= \sqrt{-{\rm det}(g_{mn} + \cF_{mn})} -f_0^*C_3 - f_0^*C_1
\cF\ -{m\over2} AdA\ \ee

in agreement with the result of reference \cite{Bergshoeff:1997cf}.


\section{Conclusions}


In this paper we have provided the full details of the derivation of the
M2-brane dynamics, including the equations of motion in a supergravity
background in component form, in the framework of superembedding. We have
used the brane action principle to derive the Green-Schwarz action for
the supermembrane starting from a closed 4-form on the worldvolume, and we
have shown how one can determine the worldvolume supervielbein and give
expressions for the bosonic worldvolume dreibein and the worldvolume
gravitino as composite fields. We have also studied the structure of the
induced $d=3, N=8$ supergeometry and show that it coincides with that of
off-shell $d=3, N=8$ conformal supergravity.

In section 6 we have introduced the notion of gauged superembedding
and applied it to $U(1)$ gauging of the supermembrane in a suitable target
space. We have shown that it leads directly to the D2-brane in Type IIA
superspace without the need to dualise the additional scalar coming from
the eleventh direction. We have done this at the level of the equations
of motion via the natural gauged version of the embedding constraint; we
have also shown how  the D2-brane action can be derived using an
appropriate modification of the construction of the Wess-Zumino term in
gauged sigma models.

Finally, we have studied the massive D2-brane starting from
$D=11$ and we have described the $D=11$ supergeometry corresponding to
the lifted massive type IIA supergravity. While this is not truly
eleven-dimensional, since it relies on the presence of a Killing vector,
it has facilitated the construction of the brane theory via gauged
superembedding. As in the massless case we have shown that the gauged
superembedding formalism leads directly to the equations of motion of the
D2-brane, and we have constructed the action which includes a
Chern-Simons term for the Born-Infeld gauge field proportional to the
mass.

\bigskip


\section*{Acknowledgments}

\bigskip

We would like to than Per Sundell and Ulf Lindstr\"om for useful
discussions. P.S.H. thanks PPARC for support under grant no
PPA/G/O/2002/00475. The work of E.S. is supported in part by NSF grant
PHY-0314712.

\begin{appendix}


\section{Conventions}


The following index conventions are used: plain (underlined) indices
refer to worldvolume and target space quantities respectively and
primed indices refer to normal spaces; indices from the beginning of
the alphabet refer to preferred bases, indices from the middle of the
alphabet to coordinate bases; Latin (Greek) indices are used for even
(odd) indices, while capital indices run over the whole space. Thus a
worldvolume preferred basis index could be $A=(a,\a)$ while a target
space coordinate index could be $\unM=(\unm,\um)$. Tensor quantities
with indices are not underlined, but if the indices are omitted or if a
target space tensor is projected on some of its indices we underline
the tensor.

In sections 6 and 7 we use plain indices for the worldvolume,
underlined indices for ten-dimensional IIA superspace and underlined
hatted indices for eleven-dimensional superspace.

For the spinor indices a two step notation is used. Initially the
target space index $\ua$, running from 1 to 32 is split in two,
$\ua\rightarrow(\a,\a')$, where both the worldvolume index $\a$ and the
normal index $\a'$ run from 1 to 16. For some purposes this is adequate
but sometimes one wishes to recognise explicitly that these indices
transform under the spin group in $d=p+1$ dimensions times an internal
symmetry group. For the membrane we set

\beqa
\psi^{\a}&\rightarrow&  \psi^{\a i}\nonumber\\
\psi^{\a'}&\rightarrow& \psi^{\a i'} \eeqa

where the spinor index on the right takes on two values while $i$ and
$i'$ both  run from 1 to 8.

We use space-favoured metrics throughout, $\h_{ab}=(-1,+1,\ldots,+1)$,
and the spacetime $\e$-tensors all have $\e^{0123\ldots}=+1$.

We use the following representation of the $D=11$ $\C$-matrices:

\beqa \C^a&=&\c^a\otimes \c_9\nonumber \w2 \C^{a'}&=& 1\otimes \c^{a'}
\eeqa

where the $d=3$ $\c$-matrices are $2\xz 2$ and the $d'=8$ $\c$-matrices
are $16\xz 16$. The charge conjugation matrix is

\beq C=\e\otimes\c_9 \eeq

In indices

\beqa
(\C^a)_{\ua}{}^{\ub}&=&(\c^a)_{\a}{}^{\b}\left(\barr{cc} \d_i{}^j & 0\\
0 & -\d_{i'}{}^{j'}\earr\right) \nonumber\w4 (\C^{a'})_{\ua}{}^{\ub}&=&
\d_{\a}{}^{\b}\left(\barr{cc} 0
&(\s^{a'})_{i}{}^{j'}\\
(\tilde\s^{a'})_{i'}{}^{j}&0\earr\right) \eeqa

and

\beq C_{\ua\ub}=\e_{\a\b}\left(\barr{cc}\d_{ij} &0\\
0&-\d_{i'j'}\earr\right) \eeq

where the $8\xz 8$ $\s$-matrices are related to the eight-dimensional
$\c$-matrices by

\beq
\c^{a'}=\left(\barr{cc}0&(\s^{a'})_{ij'}\\
(\tilde\s^{\a'})_{i'j} &0\earr\right) \eeq

Eleven-dimensional spinor indices are raised or lowered according to
the rule

\beq \psi^{\ua}=C^{\ua\ub} \psi_{\ub}\ \leftrightarrow\
\psi_{\ua}=\psi^{\ub}C_{\ub\ua} \eeq

where $C$ with upper indices is the same matrix as $C$ with lower
indices. Three-dimensional spinor indices are similarly raised or
lowered using $\e$. Eight-dimensional indices, whether spinor or
vector, are raised and lowered with the standard Euclidean metric. The
$\C$-matrices with lowered indices are

\beqa
(\C^a)_{\ua\ub}&=&(\c^a)_{\a\b}\left(\barr{cc} \d_{ij} & 0\\
0 & \d_{i'j'}\earr\right) \nonumber\w4
(\C^{a'})_{\ua\ub}&=& \e_{\a\b}\left(\barr{cc} 0&-(\s^{a'})_{ij'}\\
(\tilde\s^{a'})_{i'j}&0\earr\right) \eeqa

It is straightforward to decompose any of the eleven-dimensional
$\C$-matrices with multi-vector indices in this way. We give the
two-index $\C$-matrices as they are used most in the text:

\beqa
(\C^{ab})_{\ua\ub}&=&(\c^{ab})_{\a\b}\left(\barr{cc} \d_{ij} & 0\\
0 & \d_{i'j'}\earr\right) \nonumber\w4
(\C^{ab'})_{\ua\ub}&=&-(\c^a)_{\a\b}\left(\barr{cc} 0&-(\s^{b'})_{ij'}\\
(\tilde\s^{b'})_{i'j}&0\earr\right)\nonumber\w4
(\C^{a'b'})_{\ua\ub}&=& \e_{\a\b}\left(\barr{cc} (\s^{a'b'})_{ij'}&0\\
0&-(\tilde\s^{a'b'})_{i'j}\earr\right) \eeqa

The three-dimensional $\c$-matrices are real, symmetric with lowered
indices, and satisfy

\beq \c^a\c^b=\h^{ab} + \c^{ab} \eeq

where \beq \c^{ab}=\e^{abc}\c_c\ \leftrightarrow
\c_a=-{1\over2}\e_{abc}\c^{bc} \eeq

For the eight-dimensional $\s$-matrices one may take \beqa
\s^{a'}&=&(1,i\t_r)\nonumber\w2 \tilde\s^{a'}&=&(1,-i\t_r) \eeqa

where $r=1,\ldots 7$, and where the $\t_r$ are seven-dimensional Dirac
matrices which are purely imaginary and antisymmetric. The matrices
$\s^{a'b'c'd'}$ and $\tilde\s^{a'b'c'd'}$ are symmetric, the former
being self-dual, the latter anti-self-dual, $\s^{a'b'}$ and
$\tilde\s^{a'b'}$ are antisymmetric.


\section{Riemannian embeddings}


We briefly review some elements of the theory of Riemannian embeddings.
In this section $M$ is an $n$-dimensional manifold embedded in an
$\unn$-dimensional Riemannian manifold $(\unM,\ung)$. We recall that
the main features of an embedding are: the determination of the induced
tensors on $M$ and the splitting of the tangent bundle $\unT$
(restricted to $M$) into tangent and normal subbundles $T$ and $T'$;
the Gauss-Weingarten equations which define the induced connections on
$T$ and $T'$ and which introduce the second fundamental form; the
torsion equations which relate the target space torsion tensor to the
induced worldvolume torsion, and the Gauss-Codazzi equations which
express the curvature tensors of $T$ and $T'$ in terms of the target
space curvature and the second fundamental form.

In the case of Riemannian embeddings the metric on $\unM$ induces a
metric on $M$ in the obvious way, and we can furthermore demand that
$T'$ and $T$ are orthogonal. This gives rise to a fibre metric on $T'$.
We let $(e_a)$ be a basis of vectors on $M$, $(e_{\una})$ a basis of
vectors on $\unM$ and $(e_{a'})$ a basis for the normal bundle (dim
$T'=n'$). We have

\beq e_a=e_a{}^{\una} e_{\una} \eeq and \beq e_{a'}=e_{a'}{}^{\una}
e_{\una}. \eeq

We shall assume that the basis $(e_{\una})$ is orthonormal. We can
define a natural induced Riemannian structure on $M$ by taking the
basis $(e_a)$ to be orthonormal,

\beq \ung(e_a,e_b)=\d_{ab} \eeq

We take the normal space to be orthogonal to the tangent space to $M$,
and the basis $(e_{a'})$ to be orthonormal with respect to $\ung$ as
well. Thus

\beqa
\ung(e_a,e_{b'})&=& 0\nno \\
\ung(e_{a'},e_{b'})&=& \d_{a'b'} \eeqa

Clearly the above induces an $O(n)$ structure on $M$ and an $O(n')$
structure on $T'$.

We can interpret the embedding matrix $e_a{}^{\una}$ as a section of
$T^*\otimes\unT$ and the normal matrix $e_{a'}{}^{\una}$ as a section
of $T'^*\otimes\unT$. We define covariant derivatives of these tensors,
both denoted by $\nab$ since it should be clear from the context which
one is meant, by

\beqa \nab_a e_b{}^{\unc} &=& e_a e_b{}^{\unc}-\o_{a,b}{}^c
e_c{}^{\unc} +\o_{a,\unb}{}^{\unc} u_b{}^{\unb} \nonumber\w2 \nab_a
e_{b'}{}^{\unc} &=& e_a e_{b}'{}^{\unc}-\o_{a,b'}{}^{c'}
e_{c'}{}^{\unc} +\o_{a,\unb}{}^{\unc} e_{b'}{}^{\unb} \eeqa

where $\o_{a,\unb}{}^{\unc}$ is the target space connection pulled back
onto the world surface, $\o_{a,b}{}^c$ is a connection for $T$ and
$\o_{a,b'}{}^{c'}$ is a connection for $T'$. These connections are not
as yet specified in terms of the target space connection. The
Gauss-Weingarten equations, which do just this, can be written

\beqa \nab_a e_b{}^{\unc} &=& K_{a,b}{}^{c'}e_{c'}{}^{\unc}\nno\w2
\nab_a e_{b'}{}^{\unc} &=& K_{a,b'}{}^{c}e_{c}{}^{\unc}\ \la{gw} \eeqa

where $K_{a,b}{}^{c'}$ is the second fundamental form of the surface
and $K_{a,b'}{}^{c}$ is related to it via equation \eq{ksym} below. It
is easy to show that equations \eq{gw} are indeed the Gauss-Weingarten
equations. To see this, we note that they imply

\beq \nab_a (Y^b e_b{}^{\unc})=(\nab_a Y^b)e_b{}^{\unc}+ Y^b
K_{a,b}{}^{c'}e_{c'}{}^{\unc} \la{Y1} \eeq

and

\beq \nab_a (Y^{b'} e_{b'}{}^{\unc})=(\nab_a Y^{b'})e_{b}'{}^{\unc}+
Y^{b'} K_{a,b'}{}^{c}e_{c}{}^{\unc} \la{Y2} \eeq

where $Y$ and $Y'$ are tangential and normal vector fields
respectively. On the left-hand side of both equations the $T$-indices
are contracted out so that the covariant derivative is simply the
covariant derivative of the target space $\unM$. On the right-hand side
of \eq{Y1} the first term is tangential and defines the induced
connection for $T$ while the second term is normal and defines the
second fundamental form. Similarly, on the right-hand side of \eq{Y2}
the first term is normal and defines the induced connection on $T'$
while the second term is tangential. It is easy to show that the
connections defined by the Gauss-Weingarten equations are metric and
that

\beq K_{a,bc'}=-K_{a,c'b} \la{ksym} \eeq

One way of doing this is to define \beq K_a:=(\nab_a  e) e^{-1} \eeq

where $e$ is the $\unn\xz\unn$ matrix formed from $e_a{}^{\una}$ and
$e_{a'}{}^{\una}$. The equations

\beq K_{a,b}{}^{c}= 0 \eeq and

\beq K_{a,b'}{}^{c'}=0 \eeq

are equivalent to the Gauss-Weingarten equations above. Since $(e_a)$
and $(e_{a'})$ are orthonormal bases it follows that $e$ is an element
of $O(\unn)$ and hence that $K$ is an $\go(\unn)$-valued one-form. From
this one derives \eq{gw} immediately, while it is straightforward to
show that the induced connections  for $T$ and $T'$ are metric since
they are $\go(n)$ and $\go(n')$ valued respectively.

The torsion 2-form of $\unM$ is given by

\beq T^{\una}=d e^{\una} + e^{\unb}\wedge \o_{\unb}{}^{\una} \eeq

where $(e^{\una})$ is the basis of forms dual to $(e_{\una})$ and
$\o_{\unb}{}^{\unc}:=e^{\una}\o_{\una,\unb}{}^{\unc}$. When pulled back
onto $M$ the right-hand side becomes

\beq
 d(e^a e_a{}^{\una}) + e^a e_a{}^{\unb} \o_{\unb}{}^{\una}=
 D(e^a e_a{}^{\una})
\eeq

where $D$ is the covariant exterior derivative for the $\unT$ index.
Since the non-underlined indices are contracted one has

\beq D(e^a e_a{}^{\una})=T^a e_a{}^{\una} + e^b\wedge e^a \nab_b
e_a{}^{\una} \eeq

from which one arrives at

\beq \nab_{a} e_{b}{}^{\unc} -\nab_{b} e_{a}{}^{\unc} + T_{ab}{}^c
e_c{}^{\unc}=e_a{}^{\una} e_b{}^{\unb} T_{\una\unb}{}^{\unc}\ .
\la{tor} \eeq

Multiplying this equation by $e_{\unc}{}^{c}$ one has \beq T_{ab}{}^c=
e_a{}^{\una} e_b{}^{\unb} T_{\una\unb}{}^{\unc}e_{\unc}{}^c \eeq

using the induced connection condition. Thus the torsion of $M$ is
related to the torsion of $\unM$ in a simple fashion and clearly
vanishes if the latter does. If we multiply \eq{tor} by
$e_{\unc}{}^{c'}$ we find

\beq K_{ab}{}^{c'}-K_{ba}{}^{c'}=e_a{}^{\una} e_b{}^{\unb}
T_{\una\unb}{}^{\unc}e_{\unc}{}^{c'} \eeq

so that the second fundamental form is symmetric if the target space
torsion vanishes.

>From the definition of $K$ it is easy to derive the following relation
by differentiation

\beq \nab_a K_b-\nab_b K_a +T_{ab}{}^c K_c-[K_a,\,K_b]=(R_{ab}e)e^{-1}
\eeq

where the curvature operator acts on both indices of $e$ (with a minus
sign for the lower indices). Taking the $\go(n)$ and $\go(n')$ parts of
this equation we get the equations of Gauss and Codazzi, which are, for
the case of induced connections,

\beqa R_{ab,c}{}^{d}&=&\unR_{ab,c}{}^{d} + (K_{a,c}{}^{e'}K_{b,e'}{}^d
-(a\leftrightarrow b))\nonumber\w2
R_{ab,c'}{}^{d'}&=&\unR_{ab,c'}{}^{d'}
+(K_{a,c'}{}^{e}K_{b,e}{}^{d'}-(a\leftrightarrow b)) \eeqa

where

\beq \unR_{ab,c}{}^d=e_a{}^{\una} e_b{}^{\unb} e_c{}^{\unc}
R_{\una\unb,\unc}{}^{\und}(e^{-1})_{\und}{}^{d} \eeq

and

\beq \unR_{ab,c'}{}^{d'}=e_a{}^{\una} e_b{}^{\unb} e_{c'}{}^{\unc}
R_{\una\unb,\unc}{}^{\und}(e^{-1})_{\und}{}^{d'}. \eeq

\section{Kaluza-Klein reduction of $D=11$ superspace}

The reduction of supergravity formulated in $D=11$ superspace to ten
dimensions was outlined in \cite{Duff:1987bx}, but the full details
were not given there. Here we provide them. It is convenient to do this
in two steps: first, we make a simple reduction to what we shall refer
to as the KK frame, and then we rescale the supervielbein to go to the
string frame. The formulae specifying the KK reduction are

\bea \unE^A &=& E^A + E^{11} \chi^A\nn\w1 \unE^{11} &=& \F E^{11}
\la{KK1} \eea

where eleven-dimensional quantities are underlined and $A$ denotes a
$D=10$ type IIA superindex. The object $\chi^A$ vanishes for $A=a$,
while for $A=\a$ it denotes the spinor field of IIA supergravity. The
scalar field $\F$ is related to the dilaton in a way which will be made
precise below and the one-form $E^{11}$ is defined by

\be E^{11}=d y + C_1\ \la{KK2} \ee

where $C_1$ is the RR one-form gauge field. For the connection we put

\bea \uO_a{}^b&=&\O_a{}^b + E^{11} \O_{ya}{}^b \nn\w2
\uO_a{}^{11}&=&\O_a{}^{11} + E^{11} \O_{ya}{}^{11}\nn\w2
\uO_{\a}{}^{\b}&=&\O_{\a}{}^{\b}+{1\over2}
(\C^a\C_{11})_{\a}{}^{\b}\O_{a}{}^{11}\nn\w2 &\phantom{=}&+
E^{11}\left(\O_{y\a}{}^{\b}+{1\over2}
(\C^a\C_{11})_{\a}{}^{\b}\O_{ya}{}^{11}\right)\ . \la{KK3} \eea

These formulae define a permissible choice of the IIA connection. The
additional components of the $D=11$ connection can be read off from the
torsion. On using the above expressions in the formula for the torsion
one finds

\bea \unT_{AB}{}^c&=& T_{AB}{}^c \nn \w1 \unT_{AB}{}^\c&=& T_{AB}{}^\c
+ G_{AB} \chi^\c + \O_{[A,c}{}^{11}(\C^c\C_{11})_{B]}{}^\c \nn\w1
\unT_{AB}{}^{11}&=& \F G_{AB} +  \O_{[A,B]}{}^{11} \la{KK3.1} \eea

together with three further equations involving the additional
connection components. In the above equation quantities which are not
defined for particular values of the super indices are taken to be
zero, e.g. $\O_{A,\b}{}^{11}$, while $G=d C_1$ is the RR field strength
2-form. After some algebraic manipulation one finds

\bea T_{\a\b}{}^c   &= &-i(\C^c)_{\a\b}  \nn\w1 T_{\a b}{}^c
&=&T_{ab}{}^c=0     \nn\w1 T_{\a\b}{}^\c
&=&i\F^{-1}(\C_{11})_{\a\b}\chi^{\c} -i\F^{-1}
                  (\C^c\chi)_{(\a}(\C_c \C_{11})_{\b)}{}^{\c} \nn\w1
T_{a\b}{}^{\c}&=&\unT_{a\b}{}^{\c}-G_{a\b}\chi^{\c}+{1\over4}
G_{ab}(\C^b\C_{11})_{\b}{}^{\c} \ . \la{KK4} \eea

The components of $G_{AB}$ are also determined by the reduction
procedure. They are

\bea G_{\a\b}&=&-i\F^{-1}(\C_{11})_{\a\b}\nn\w1 G_{\a b}&=&
-i\F^{-2}(\C_b\chi)_{\a} \la{KK5} \eea

and $G_{ab}$. We also find that

\be \chi_{\a}=\nab_{\a}\F \ . \la{KK6} \ee

The additional components of the eleven-dimensional connection are
given by

\bea \O_{\a,b 11}&=&i\F^{-1}(\C_b\chi)_{\a}\nn\w1 \O_{a,b
11}&=&-{1\over2}\F G_{ab}\nn\w1 \O_{y,b 11}&=& \nab_a \F\nn\w1
\O_{y,ab}&=&{1\over2}\F^2 G_{ab}\ . \la{KK7} \eea

The components of the IIA 3-and 4-form field strengths can be read off
using the formulae

\bea G_{ABCD}&=& \unG_{ABCD}\nn\w1 H_{ABC}&=&\F\unG_{ABC 11}- \chi^\d
\unG_{\d ABC}. \la{KK8} \eea

One finds that the non-vanishing components are

\bea G_{\a\b cd}&=& -i(\C_{cd})_{\a\b}\nn\w1
G_{abcd}&=&\unG_{abcd}\nn\w1 H_{\a\b c}&=&
-i\F(\C_c\C_{11})_{\a\b}\nn\w1 H_{\a bc}&=&-i(\C_{bc}\chi)_{\a}\nn\w1
H_{abc}&=&\F\unG_{abc 11}\ . \la{KK9} \eea

The string frame can be reached by the rescalings

\be E^a\mapsto e^{\f\over3} E^a\qquad E^\a\mapsto e^{\f\over 6}E^\a
\la{KK10} \ee

provided that we set

\be \F=e^{2\f\over3} \la{KK11} \ee

where $\f$ is the dilaton. It is also necessary to redefine the bosonic
part of the connection if one wishes to maintain the standard
constraint $T_{ab}{}^c=0$, but we shall not bother to do this here. The
new components of the torsion up to dimension one-half in the string
frame are

\bea T_{\a\b}{}^c   &= &-i(\C^c)_{\a\b}  \nn\w2 T_{\a b}{}^c   &=
&\d_b{}^c \l_\a   \nn\w2 T_{\a\b}{}^\c  &=&
-2(\C_{11})_{\a\b}(\C_{11}\l)^{\c} +
                  2(\C^c\C_{11}\l)_{(\a}(\C_c \C_{11})_{\b)}{}^{\c}
                  +\d_{(\a}{}^\c\l_{\b)}
\la{KK12} \eea

where we have introduced

\be \l_\a:=\frac{1}{3} \nab_\a \f\ . \la{KK13} \ee

For the non-zero components of the form field strengths up to dimension
one-half we have

\bea G_{\a\b}&=&-ie^{-\f}(\C_{11})_{\a\b} \nn\w1 G_{\a
b}&=&2e^{-\f}(\C_b\C_{11}\l)_\a \nn\w1 G_{\a\b cd}&=&-i e^{-\f}
(\C_{cd})_{\a\b}\nn\w1 H_{\a\b c}&=&-i(\C_c\C_{11})_{\a\b}\nn\w1 H_{\a
bc}&=& 2(\C_{bc}\C_{11}\l)_{\a}\ . \la{KK14} \eea

These results seem to be compatible with the expressions given for the
superspace tensors of type IIA supergravity given in references
\cite{Carr:1986tk,Cederwall:1996ri} but we were not able to find a
redefinition which would take them into the form given in
\cite{Bergshoeff:1997cf}.

\end{appendix}

\end{document}